%% file: preprint_sicomore.tex
\documentclass{article}
\usepackage{arxiv}

\usepackage[english]{babel}
\usepackage{empheq}
\usepackage[ruled,linesnumbered,algoruled]{algorithm2e}
\usepackage{tikz}
\usepackage{lipsum}

\usepackage{subfigure}
\usepackage{natbib}
\usepackage{enumerate}
\usepackage{multirow}
\usepackage{dsfont}
\usepackage{tcolorbox}
\usepackage{upgreek}
\usepackage{placeins}
\tcbuselibrary{breakable}

\usepackage{makecell}
\usepackage{tabularx}

\usepackage{booktabs}
\usetikzlibrary{arrows, shapes, snakes, backgrounds, patterns}

\usepackage[utf8]{inputenc} 
\usepackage[T1]{fontenc}    
\usepackage{hyperref}       
\usepackage{url}            
\usepackage{booktabs}       
\usepackage{amsfonts, amsthm}       
\usepackage{nicefrac}       
\usepackage{microtype}      
\usepackage{placeins}

\input{commandes-math}
\input{commandes-dessin}

\input{commandes-colors}

\newif\iflongversion
\longversiontrue

\newif\ifnewresults
\newresultstrue

\newif\ifstabs

\ifnewresults
\stabstrue
\fi

\title{Fast Computation of Genome-Metagenome Interaction Effects}
\author{
  Florent Guinot \\
  L'Oréal, R\&D, 94550, Chevilly-Larue, France \\
  Université Paris-Saclay, CNRS, Univ Évry, Laboratoire de
  Mathématiques et Modélisation d'Évry \\ 91037, Évry-Courcouronnes,
  France \\
  \And
  Marie Szafranski  \\
  ENSIIE, 91025, Évry-Courcouronnes,
  France \\
  Université Paris-Saclay, CNRS, Univ Évry, Laboratoire de
  Mathématiques et Modélisation d'Évry \\
  91037, Évry-Courcouronnes,
  France\\
  \texttt{marie.szafranski@math.cnrs.fr} \\
  \And
  Julien Chiquet \\
  Université Paris-Saclay, AgroParisTech, INRAE, UMR MIA-Paris, 
  75005,
  Paris, France \\
  \And
  Anouk Zancarini \\
  Plant Hormone Biology, Swammerdam Institute
  for Life Sciences, University of Amsterdam \\
  1098 XH, Amsterdam, The Netherlands \\
  \And
  Christine Le Signor \\
  UMR 1347 Agro\'ecologie, AgroSup Dijon, CNRS,
  Univ. Bourgogne, INRAE, Univ. Bourgogne Franche-Comt\'e \\
  21000, Dijon, France \\
  \And
  Christophe Mougel \\
  UMR 1349 IGEPP, INRAE, Agrocampus Ouest, Univ.  Rennes 1, 
  35653, Le Rheu, France
  \And
  Christophe Ambroise \\
  Université Paris-Saclay, CNRS, Univ Évry, Laboratoire de
  Mathématiques et Modélisation d'Évry \\
  91037, Évry-Courcouronnes, France 
}

\begin{document}

\maketitle

\begin{abstract}
   \textbf{Motivation.} Association studies have been widely used to
  search for associations between common genetic variants observations
  and a given phenotype. However, it is now generally accepted that
  genes and environment must be examined jointly when estimating
  phenotypic variance.  In this work we consider two types of
  biological markers: genotypic markers, which characterize an
  observation in terms of inherited genetic information, and
  metagenomic marker which are related to the environment.  Both types
  of markers are available in their millions and can be used to
  characterize any observation uniquely.
  
  \textbf{Objective.} Our focus is on detecting interactions between
  groups of genetic and metagenomic markers in order to gain a better
  understanding of the complex relationship between environment and
  genome in the expression of a given phenotype.

  \textbf{Contributions.} We propose a novel approach for
  efficiently detecting interactions between complementary datasets in
  a high-dimensional setting with a reduced computational cost.  The
  method, named SICOMORE, reduces the dimension of the search space by
  selecting a subset of supervariables in the two complementary
  datasets. These supervariables are given by a weighted group
  structure defined on sets of variables at different scales. A Lasso
  selection is then applied on each type of supervariable to obtain a
  subset of potential interactions that will be explored via linear
  model testing.

  \textbf{Results.} We compare SICOMORE with other approaches in
  simulations, with varying sample sizes, noise, and numbers of true
  interactions. SICOMORE exhibits convincing results in terms of
  recall, as well as competitive performances with respect to running
  time. The method is also used to detect interaction between genomic
  markers in \textit{Medicago truncatula} and metagenomic markers in
  its rhizosphere bacterial community.

  \textbf{Software availability.} A \texttt{R} package is
  available~\citep{ACGS:pkg_SICOMORE:2020}, along with its
  documentation and associated scripts, allowing the reader to
  reproduce the results presented in the paper.
\end{abstract}



\section{Introduction}
\label{sec:intro}

{Association studies are a popular approach for digging out genetic
  information relating to a given phenotype. To avoid confusion
  effects (e.g. stratification due to population origin) and improve
  the diagnostic, it is common practice to integrate environmental
  data in the analysis. These additional variables are generally few
  in number, of the order of ten.}

{In this paper we propose a generic method for taking thousands or
  even millions of environmental variables into consideration, with
  the aim of finding significant interactions between these variables
  and genetic markers. We illustrate the proposed algorithm on the
  genome of \textit{Medicago truncatula (Fabaceae, Plantae)} and
  metagenomic markers in its rhizosphere bacterial community, but it
  could be applied in many other contexts.}

\subsection{Gene-environment interactions}

Genome-Wide Association Studies (GWAS) look for genetic markers linked
to a phenotype of interest. Typically, hundreds of thousands of single
nucleotide polymorphisms (SNPs) are analyzed with a limited sample
size using high-density genotyping arrays. GWAS are a powerful tool
for investigating the genetic architecture of complex biological
processes and have been successful in identifying hundreds of
associated variants. However, they have been able to explain only a
small proportion of the phenotypic variations expected from classical
linkage analysis~\citep{Manolio2009Finding}.

Some of the missing heritability may be uncovered by taking into
account correlations among variables and epistasis \citep[and
references therein]{stanislas2016eigen}. %
{Another way to understand and improve the knowledge of complex
  phenotypes is to look at gene-environment interactions. If the
  contributions of genes and environment to a phenotype are examined
  separately and interactions between them ignored, this can give
  incorrect estimates of how much phenotypic variance is attributable
  to genes alone, to environment alone, and to genes and environment
  jointly.}

Gene-environment interactions are clearly of great interest in medical
genetics and epidemiology~\citep{clavel2007progress, thomas2010gene}
{but also in plant research regarding environmental adaptation
  issues~\citep{hancock2011adaptation, hassani2018microbial}}.
{In particular, Metagenome-Wide Association Analysis
  (MWAS)~\citep{segata2011metagenomic, wang_2107_human,
    wang2016metagenome} is providing a growing body of evidence
  regarding the role of gut microbiome in basic biological processes
  and in the development and progression of major human diseases, such
  as infectious diseases, gastrointestinal cancers, and metabolic
  diseases.}
{In plants, the role of rhizosphere\footnote{{The rhizosphere was
      defined by Hiltner in 1904 as the area around a plant root that
      is inhabited by a unique population of microorganisms influenced
      by the chemicals released from plant roots.}}  microbiome on the
  plant growth and health is well known and has been studied since the
  early 2000s~\citep{mukerji2002techniques, pinton2007rhizosphere,
    lugtenberg2009plant, berendsen2012rhizosphere}.}
{While GWAS analyses have been able to identify associations between
  the plant genome of \textit{Arabidopsis thaliana} and the metagenome
  (amplicon sequencing) of its associated phyllosphere and root
  microbial communities~\citep{horton2014genome,
    bergelson2019characterizing}, in plants, to our knowledge, no
  specific MWAS analyses have so far been done.}

\subsection{Combining genome and metagenome analyses}

{There have been a number of works regarding the integration of
  multi-omics data in statistical or machine learning models, with
  several review papers. For instance, \cite{Li2016briefing}
  {establish a typology regarding different families of models.}
  \cite{huang2017frontiers} also list the kind of omics data which can
  be used and the outputs given by the methods.
  \cite{hawe2019frontiers} pay attention to the inference of
  interaction networks.}

{However, these methods do not include environmental variables and
  consequently fail to address specificities of such features. There
  exists literature discussing both microbiome and genetics. They are
  mainly classical methods applied to a reduced set of species-gene
  pairs~\citep{knights2014complex}.}
Another way of relating genetic and metagenomic data is to consider
the metagenome as a phenotype and to perform quantitative trait locus
(QTL) mapping.  This kind of metagenomic QTL analysis illustrates the
role of host genetics in shaping metagenomic diversity between
individuals~\citep{srinivas2013genome, wang2016genome}.

An alternative of interest is to consider metagenomic variables as
environmental variables in GWAS. %
{Several quantitative approaches have been proposed in classical
  gene-environment interaction studies with a small number of
  environmental factors limited to certain modalities, such as
  different status {(smoking / non smoking, for instance)} or
  {medical} treatments~\citep{hutter2013gene, han2018review}.} %
{More specifically, our proposal shares similarities with approaches
  where} interactions can be modelled 
using a classical (generalized) linear model with interaction terms
\citep{lin2013interactionsGLM}.

{However, the number of interactions that need to be tested may
  increase dramatically when metagenomic markers are considered as
  environmental data.} In this perspective, variable selection or
variable compression may be of use here as a means of reducing the
dimension of the problem {in order to design an efficient method for
  detecting gene-environment interaction in a high-dimensional
  setting}.

\subsection{Taking structures into account in association studies}

Data compression for dimension reduction may be achieved in various
ways.  A distinction is usually drawn between feature selection and
feature extraction. Feature selection consists in selecting a few
relevant variables from among the original variables, whereas feature
extraction consists in computing new representative variables.

For the kind of association study that concerns us here, feature
selection is often preferred to feature extraction for interpretative
purposes. In this paper we advocate a mixed approach including feature
extraction that is based on the underlying structures of genome and
metagenome, combined with feature selection.

The idea of considering group structures is not new. It has already
been advocated both in the context of GWAS
\citep{dehman2015performance} and MWAS \citep{qin2012metagenome}.
In the context of prediction from gene expression regression,
\cite{PHT:Biostat07} proposed clustering genes hierarchically to
obtain a dendrogram that reveals their nested correlation
structure. At each level of the hierarchy, supergenes are computed as
the average expression of the current clusters.  It can be shown that
regressing over supergenes improves precision if the correlation
structure is sufficiently strong. In a similar fashion,
\cite{guinot2017learning} made use of the haplotype structure of the
human genome when they proposed a dimension-reduction approach that
can be applied in the context of GWAS.
It is worth noting that similar ideas have also been developed in
other areas such medical imaging~\citep{CST:ArXiv18}.

\subsection{Contributions and organization of the paper}

{In this work, we propose a 
  method for detecting interactions between genomic and metagenomic
  data.  The method comprises four steps. Given a dataset:
\begin{enumerate}[(1)]
\item Identify a group structure within the variables using a
  hierarchical clustering;
\item Create compressed features, or \emph{supervariables}, according
  to this group structure;
\item Select a subset of supervariables using a Lasso procedure with a
  penalty factor weighted by the length of the gap between two
  successive levels of a hierarchical clustering;
\item Combine the two compressed datasets in a linear model with
  interactions in order to perform multiple hypothesis testing.
\end{enumerate}
This scheme allows interactions to be detected efficiently in a
high-dimensional setting with a reduced computational cost.}

The paper is organized as follows. Section~\ref{sec:model} looks at
the role of linear models of interactions and proposes a framework for
learning using complementary datasets.
Section~\ref{sec:implementation} describes our method, which seeks to
uncover relevant interactions using, first, compressions of data based
on hierarchical structures, second, a Lasso selection procedure and,
third, model testing.  Finally, Section~\ref{sec:XP_simu} provides an
illustration of our approach using numerical simulations, and
Section~\ref{sec:XP_INRA} describes an application for examining
interactions between the 
{genomic markers} of the species \textit{Medicago truncatula} and
{metagenomic markers} of its rhizosphere microbial community.

\section{Learning interactions with complementary datasets}
\label{sec:model}

This section gives a general introduction together with some notation,
and outlines how we will establish a compact model of interactions
between complementary datasets.

\begin{rem}
  {Here, and in what follows, the term \emph{genomic} data will refer
    to SNP data. In Sections~\ref{sec:model},~\ref{sec:implementation}
    and~\ref{sec:XP_simu}, we will use 
    the term \emph{metagenomic} data for
    metabarcoding or shotgun data. The application on \textit{Medicago
      truncatula} will be described in greater detail. Extensions to
    other kinds of data will be discussed in
    Section~\ref{sec:conclusion}.}
\end{rem}

\subsection{Setting and notations}

Let us consider observations from two complementary views, $\vu$ (for
Genomic data) and $\vd$ (for Metagenomic data), which are placed
together in a training set
$\mathcal{S} = \{(\bx^\vu_i, \bx^\vd_i, y_i)\}_{i=1}^N$, where
$(\bx^\vu_i, \bx^\vd_i, y_i) \in \IR^{D_\vu} \times \IR^{D_\vd} \times
\IR$.

We assume the existence of underlying biological information on $\vu$
and $\vd$, encoded as groups.  The group structure over $\vu$ is
defined by $N_{\vu}$ groups of variables
$\gvu=\{\gvu_{\ivu} \}_{\ivu=1}^{N_{\vu}}$.  We denote as
$\bx_i^{\ivu} \in \IR^{D_{\ivu}}$ the sample $i$ restricted to the
variables of $\vu$ from group $\gvu_{\ivu}$. Similarly, the group
structure over $\vd$ is defined by $N_{\vd}$ groups of variables
$\gvd=\{\gvd_{\ivd}\}_{\ivd=1}^{N_{\vd}}$, and
$\bx_i^{\ivd} \in \IR^{D_{\ivd}}$ is the sample $i$ restricted to the
variables of $\vd$ from group $\gvd_{\ivd}$.

We also introduce $D_I = D_{\vu} \cdot D_{\vd}$ and
$N_I = N_{\vu} \cdot N_{\vd}$, corresponding to the number of
variables and the number of groups that may interact.

Finally, we use the following convention: vectors of observations
indexed with $i$, such as $\bx_i$, will usually be row vectors, while
vectors of coefficients, such as $\bbeta$, will usually be column
vectors.

\subsection{Interactions in linear models}

Interactions between data from views $\vu$ and $\vd$ may be
captured in the model
\begin{equation}
  \label{eq:classical_interaction_model}
  y_i =   \bx^{\vu}_i \bgamma_{\vu}
  + \bx^{\vd}_i \bgamma_{\vd}
  + \bx^{\vu}_i \bDelta_{\vu\vd} (\bx^{\vd}_i)^T  +
  \epsilon_i \,,
\end{equation}
where the vectors $\bgamma_{\vu} \in \IR^{D_{\vu}}$ and
$\bgamma_{\vd} \in \IR^{D_{\vd}}$ denote the linear
effects related to $\vu$ and $\vd$ respectively, the matrix
$\bDelta_{\vu\vd} \in \IR^{D_{\vu} \times D_{\vd}}$ contains the
interactions between all pairs of variables in $\vu$ and $\vd$, and
$\epsilon_i \in \IR$ is a residual error.

Models with interactions distinguish between \emph{strong dependency}
(SD) and \emph{weak dependency} (WD). \emph{Strong dependency} is the
more common hypothesis (see for instance \citep{bien2013Lasso} and the
discussion therein), and it means that an interaction is effective if
and only if the corresponding single effects are also
effective. \emph{Weak dependency}, on the other hand, means that an
interaction is effective if
one of the main effects is also effective. Formally, for all variables
$j \in \bx^{\vu}$ and for all variables $j' \in \bx^{\vd}$, if
$\gamma_{j}$, $\gamma_{j'}$ and $\delta_{jj'}$ are the coefficients
related to $\bgamma_{\vu}$, $\bgamma_{\vd}$ and $\bDelta_{\vu\vd}$,
then
\begin{align*}
  & (SD) && \emph{} \delta_{jj'} \neq 0 \qquad \Rightarrow  \qquad
            \gamma_{j}
            \neq 0 && \text{ and } && \gamma_{j'} \neq 0  \,, \\
  &  (WD) && \emph{} \delta_{jj'} \neq 0 \qquad \Rightarrow  \qquad
             \gamma_{j}
             \neq 0 && \text{ or } && \gamma_{j'} \neq 0  \,.
\end{align*}
In this context, \cite{bien2013Lasso} proposed a sparse model of
interactions that is likely to encounter computational limitations for
large-dimensional problems (\cite{lim2015learning} and
\cite{she2016group}). \cite{lim2015learning} present a method for
learning pairwise interactions in a regression model by solving a
constrained overlapping group Lasso \citep{JOV:ICML09} in a manner
that satisfies strong dependencies. \cite{she2016group} propose a
formulation with an overlapping regularization that fits both types of
hypothesis, and they provide theoretical insights on the resulting
estimators.~\footnote{To our knowledge, their implementation based on
  an alternating direction method of multipliers is not publicly
  available.}

However, the dimension $D_{\vu} + D_{\vd} + D_I$ inherent in
Problem~\eqref{eq:classical_interaction_model} when estimating
$\bgamma_{\vu}$, $\bgamma_{\vd}$ and $\bDelta_{\vu\vd}$ may be
inconveniently large, especially for applications with numerous 
variables such as in biology with genomic and metagenomic markers.  To
reduce this dimension we propose compressing the data according to an
underlying structure that may be defined on the basis of prior
knowledge or uncovered using clustering algorithms.

\subsection{Compact model}
\label{sec:compress-data}

Let us consider that if we have a compression function for all groups
$\vu$ and $\vd$, we can shape
Problem~\eqref{eq:classical_interaction_model} into a compact form
\begin{equation}
  \label{eq:compact_detailled_model}
  y_i =
  \sum_{\ivu \in \gvu} \tilde{x}_i^\ivu \beta_{\ivu} +
  \sum_{\ivd \in \gvd} \tilde{x}_i^\ivd \beta_{\ivd} +
  \sum_{\ivu \in \gvu} \sum_{\ivd \in \gvd}
  \underbrace{\left(\tilde{x}_i^\ivu  \cdot
      \tilde{x}_i^\ivd\right)}_{\phi^{\ivu\ivd}_i}
  {\theta_{\ivu\ivd}} + \, \epsilon_i \,,
\end{equation}
where $\tilde{x}_i^\ivu \in \IR$ is the $i^{th}$ compressed sample of
the variables that belong to the group $\ivu$ for the view $\vu$, and
$\beta_{\ivu} \in \IR$ is its corresponding coefficient. The
counterparts in the group $\ivd$ for the view $\vd$ are
$\tilde{x}_i^\ivd \in \IR$ and $\beta_{\ivd} \in \IR$.  Finally,
$\theta_{\ivu\ivd} \in \IR$ is the interaction between groups $\ivu$
and $\ivd$.

Problem~\eqref{eq:compact_detailled_model} can be reformulated in a
vector form. Let $\tbx_{i} \in \IR^{N_{\vu}}$,
$\bbeta_{\vu} \in \IR^{N_{\vu}}$, $\tbx_{i} \in \IR^{N_{\vd}}$ and
$\bbeta_{\vd} \in \IR^{N_{\vd}}$ be
\begin{align*}
  \tbx_{i}^{\vu} & = (\tilde{x}_i^1 \cdots \tilde{x}_i^\ivu \cdots
                   \tilde{x}_i^{N_{\vu}})\,,
  &  \bbeta_{\vu} & = (\beta_{1} \cdots \beta_{\ivu}
                    \cdots \beta_{N_{\vu}})^{T}\,,\\
  \tbx_{i}^{\vd} & = (\tilde{x}_i^1 \cdots \tilde{x}_i^\ivd \cdots
                   \tilde{x}_i^{N_{\vd}}) \,,
  & \bbeta_{\vd} & = (\beta_{1} \cdots \beta_{\ivd}
                   \cdots \beta_{N_{\vd}})^{T}\,.
\end{align*}

We denote as $\bphi_{i} \in \IR^{N_I}$ the vector whose general
component is given by $\phi^{\ivu\ivd}_i$ in
Equation~\eqref{eq:compact_detailled_model}, that is
\begin{align*}
  \bphi_{i} & =
                       \left(
                       \phi^{11}_i \cdots
                       \phi^{1N_{\vd}}_i
                       \cdots
                       \phi^{\ivu \ivd}_i
                       \cdots
                       \phi^{{N_{\vu}} 1}_i
                       \cdots
                       \phi^{{N_{\vu}}  {N_{\vd}}}_i
                       \right) \,,
\end{align*}
and $\btheta \in \IR^{N_I}$ denotes the corresponding
vector of coefficients, that is
\begin{align*}
  \btheta = & \left(\theta_{11}   \cdots \theta_{1
              {N_{\vd}}}  \cdots \theta_{\ivu \ivd}
              \cdots \theta_{{N_{\vu}} 1}  \cdots
              \theta_{{N_{\vu}}  {N_{\vd}}} \right)^T\,.
\end{align*}

Finally, Problem~\eqref{eq:compact_detailled_model}
reads as a classical linear regression problem
\begin{equation}
  \label{eq:compact_vector_model}
  y_i =  \tbx_{i}^{\vu} \bbeta_{\vu}  +
  \tbx_{i}^{\vd} \bbeta_{\vd} +
  \bphi_{i}  \btheta  +
  \epsilon_i \,,
\end{equation}
of dimension  $N_{\vu}  + N_{\vd} + N_I$.

\subsection{Uncovering relevant interactions}
\label{sec:recover-interactions}

Compared to Problem~\eqref{eq:classical_interaction_model} and
provided that $N_\vu$ and $N_\vd$ are reasonably smaller than
$D_{\vu}$ and $D_{\vd}$, the dimension of
Problem~\eqref{eq:compact_vector_model} is drastically reduced, so
that it may be solved with the aid of a suitable optimization
algorithm and sufficient computing resources. For instance,
\cite{DT:IEEEIT2008} give an overview of $\ell_1$ regularized
algorithms to solve sparse problems like Lasso, which in our case
could take the form:
\begin{empheq}[left={\empheqlbrace}]{align*}
   \label{eq:Lasso_interaction}
  \argmin_{\bbeta_{\vu},\, \bbeta_{\vd},\, \btheta}
  & \quad \sum_{i=1}^n \left(y_i - \tbx_{i}^{\vu} \bbeta_{\vu} -
    \tbx_{i}^{\vd}  \bbeta_{\vd}  - \bphi_{i}  \btheta \right)^2 \\
  &  \quad + \lambda_\vu \sum_{\ivu=1}^{N_{\vu}} |\beta_{\ivu}|
  +  \lambda_\vd \sum_{\ivd=1}^{N_{\vd}} |\beta_{\ivd}|
  +  \lambda_I
  \sum_{\ivu,\ivd=1}^{N_I}  |\theta_{\ivu\ivd}| \,,
\end{empheq}
with $\lambda_\vu$, $\lambda_\vd$ and $\lambda_I$ being the positive
hyperparameters that respectively control the amount of sparsity
related to coefficients $\bbeta_\vu$, $\bbeta_\vd$ and $\btheta$. The
$N_\vu + N_\vd + N_I$ dimension may nevertheless remain large in
relation to the number of observations $N$. Also, it will be remarked
that this kind of formulation does not automatically entail the
dependency hypotheses (SD) and (WD) unless additional constraints are
introduced. For this purpose, the works by \cite{bien2013Lasso,
  lim2015learning} or \cite{she2016group} mentioned above may be
considered. In the following section we present another way of
reducing the dimension further and ensuring that the strong dependency
hypothesis is satisfied.


\section{Method}
\label{sec:implementation}

In this section we provide some elements for addressing
Problem~\eqref{eq:compact_vector_model} in relation to biological
problems involving complementary datasets.  Our proposed approach,
which we have named \textbf{SICOMORE} ({Selection of Interaction
  effects in COmpressed Multiple Omics REpresentations}), is available
for download as an \texttt{R} package~\citep{ACGS:pkg_SICOMORE:2020}.

\subsection{Preprocessing of the data}
\label{sec:preprocess}

When tackling problems that involve genomic and metagenomic
interactions, some prior transformations are necessary. This
preliminary step may also include a first attempt at reducing the
dimension.

\subsubsection*{Transformation for metagenomic data}

Metagenome sequencing gives rise to features that take the form of
proportions in different samples. This kind of information is referred
to in the statistical literature as compositional data
\citep{Aitchison:JRSSB82} and is known to be subject to negative
correlation bias \citep{Pearson:RSL1896, Aitchison:JRSSB82}. The most
common way to circumvent this issue is to transform the $D_{\vd}$
features using centered log-ratios 
and to replace $0$ values using maximum-likelihood approaches
(see \citep{GMVF:AJS16, gloor2017microbiome} and references
therein). A more detailed presentation of these aspects may be found
in~\citep{rau_2017_statistical}.

\subsubsection*{Initial selection of variables}

As described in Section \ref{sec:model}, we make the assumption that
interactions have strong dependencies, which means that an interaction
can be effective only if the two simple effects associated with the
variables in interaction are included in the model.
For this reason it may be advantageous to make an initial selection in
order to eliminate inoperative single effects on $\vu$ and $\vd$
respectively.
Different approaches for carrying out this selection may be considered.
For example, screening rules can eliminate variables that will not
contribute to the optimal solution of a sparse problem, sweeping all
the variables upstream to the optimization.  In cases where this kind
of screening is appropriate, the work of \cite{Lee:PAMI17} is a useful
resource. Their focus is on Lasso problems and they present an %
overview of these techniques, together with an ensemble of screening
rules. Once the screening has been performed, the optimization of a
Lasso problem gives the final set of variables.

\subsection{Structuring the data}
\label{sec:structure}

Once the data have been preprocessed, hierarchical
clustering using Ward's method with appropriate
distances can be employed to uncover the tree structures.

\subsubsection*{Clustering of metagenomic data}

{Several approaches are available for analyzing {microbiota}
  compositions.  \cite{li2015microbiome} has produced a review of
  statistical and computational methods according to different
  objectives and/or technologies. For problems with numerous similar
  reference sequences, \cite{fisher2017abundance} have proposed a
  general linear model approach designed to estimate taxon abundances
  for strain-level analyses.}

A commonly used approach when analyzing metabarcoding data is to group
sequences into taxonomic units~\citep{blaxter2005defining}. The
features arising from such a sequencing are often modeled as
Operational Taxonomic Units (OTUs), each OTU representing 
species proxies according to some degree of sequence 
similarity. %
{More recent methods based on denoising techniques have led to the
  definition of Amplicon Sequence Variants (ASVs), which can be
  considered as refined versions of OTUs~\citep{callahan2017exact}.}

While the structure of microbial communities can be defined according
to the underlying phylogenetic tree, it also makes sense to use more
classical distances to define a hierarchy based on the abundance of
OTUs. In our application, we use an agglomerative hierarchical
clustering with the Ward criterion.

\subsubsection*{Clustering of genomic data}

When the genomic information is available through SNP, the tree
structure on $\vu$ will be defined using a hierarchical clustering
algorithm that integrates the linkage disequilibrium as the measure of
dissimilarity~\citep{dehman2015performance}.

{This algorithm is a computationally efficient hierarchical clustering
  that makes use of the structure of the genome in order to cluster
  SNPs into adjacent groups. More specifically, it is a spatially
  constrained hierarchical clustering based on Ward's incremental
  sum-of-squares algorithm~\citep{ward1963hierarchical} in which the
  measure of dissimilarity is based on the linkage disequilibrium
  between SNPs $j$ and $j'$: \(1 - r^2(j,j')\). The algorithm also
  makes use of the fact that the linkage disequilibrium matrix can be
  modeled as block-diagonal by allowing only groups of variables that
  are adjacent on the genome to be merged, which significantly reduces
  the computational cost.}

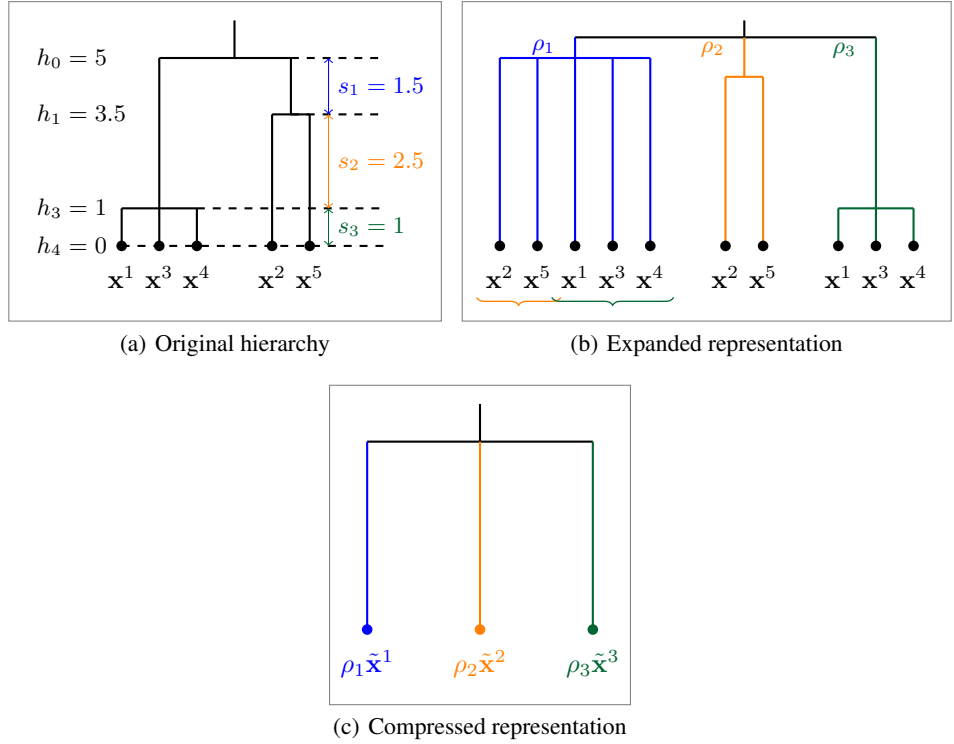
\begin{figure*}[th]
  \tikzstyle{background grid}=[thin, draw=gray, step=0.5cm]
  \centering 
  \subfigure[Original hierarchy]{%
    \label{fig:p-h:original}
    \begin{tikzpicture}
      \draw[color = bgfig, draw=gray] (-1, -0.5) rectangle
      (4.75, 3.75);

      \node at (-0.75,3) [right]{\small$h_0 = 5$};
      \node at (-0.75,2.25) [right]{\small$h_1 = 3.5$};
      \node at (-0.75,1) [right]{\small$h_3 = 1$};
      \node at (-0.75,0.5) [right]{\small$h_4 = 0$};
      \fill[black] (0.5, 0.5) circle (2pt) node (x1)
      [black,below=0.15cm] {$\small \bx^1$}; 
      \fill[black]  (1, 0.5) circle (2pt) node (x3)
      [black,below=0.15cm] {$\small \bx^3$}; 
      \fill[black]  (1.5, 0.5) circle (2pt) node (x4)
      [black,below=0.15cm] {$\small \bx^4$}; 

      \draw[-, thick] (0.5, 1) -- (1.5, 1);      
      \draw[-, thick] (0.5, 1) -- (0.5, 0.5);
      \draw[-, thick] (1, 1) -- (1, 0.5);
      \draw[-, thick] (1.5, 1) -- (1.5, 0.5);

      \fill[black]  (2.5, 0.5) circle (2pt) node (x2)
      [black,below=0.15cm] {$\small \bx^2$}; 
      \fill[black]  (3, 0.5) circle (2pt) node (x5)
      [black,below=0.15cm] {$\small \bx^5$};

      \draw[-, thick] (2.5, 2.25) -- (3, 2.25);      
      \draw[-, thick] (2.5, 2.25) -- (2.5, 0.5);
      \draw[-, thick] (3, 2.25) -- (3, 0.5);

      \draw[-, thick] (2, 3) -- (2, 3.5);   
      \draw[-, thick] (1, 3) -- (2.75, 3);   
      \draw[-, thick] (1, 3) -- (1, 1);   
      \draw[-, thick] (2.75, 3) -- (2.75, 2.25);   

      \draw[dashed, thick] (2.75, 3) -- (4, 3);   
      \draw[dashed, thick] (2.75, 2.25) -- (4, 2.25);  

      \draw[<->, blue] (3.25, 3) -- (3.25, 2.25) node [midway,
      right]{\small $s_1=1.5$};

      \draw[dashed, thick] (1.5, 1) -- (4, 1);  

      \draw[<->, orange] (3.25, 2.25) -- (3.25, 1)node [midway,
      right]{\small $s_2=2.5$};

      \draw[dashed, thick] (0.5, 0.5) -- (4, 0.5);  

      \draw[<->, darkgreen] (3.25, 1) -- (3.25, 0.5)node [midway,
      right]{\small $s_3=1$};

    \end{tikzpicture}
  }
  \subfigure[Expanded representation]{
    \label{fig:p-h:expanded}
    \begin{tikzpicture}
      \draw[color = bgfig, draw=gray] (0, -0.5) rectangle
      (6.5, 3.75);

      \draw[-, thick] (3.75, 3.275) -- (3.75, 3.5);
      \draw[-, thick] (1.5, 3.275) -- (5.5, 3.275);
      \draw[-, thick, blue] (1.5, 3.275) -- (1.5, 3) node [pos=0.5,
      label = left:{\small$\textcolor{blue}{\rho_1}$}]{};

      \draw[-, thick, orange] (3.75, 3.275) -- (3.75, 2.75) node [pos
      = 0.3, label = left:{\small$\textcolor{orange}{\rho_2}$}]{};

      \draw[-, thick, darkgreen] (5.5, 3.275) -- (5.5, 1) node [pos =
      0.075, label = left:{\small$\textcolor{darkgreen}{\rho_3}$}]{};

      \draw[-, blue, thick] (0.5, 3) -- (2.5, 3);
      \draw[-, blue, thick] (0.5, 3) -- (0.5, 0.5);
      \draw[-, blue, thick] (1, 3) -- (1, 0.5);
      \draw[-, blue, thick] (1.5, 3) -- (1.5, 0.5);
      \draw[-, blue, thick] (2, 3) -- (2, 0.5);
      \draw[-, blue, thick] (2.5, 3) -- (2.5, 0.5);

      \fill[black] (0.5, 0.5) circle (2pt) node (x2root)
      [black,below=0.15cm] {$\small \bx^2$}; 
      \fill[black]  (1, 0.5) circle (2pt) node (x5root)
      [black,below=0.15cm] {$\small \bx^5$}; 

      \draw[snake=brace, mirror snake, color = orange, segment
      amplitude=3pt, line width=0.5pt] (x2root.south west) --
      (x5root.south east);

      \fill[black]  (1.5, 0.5) circle (2pt) node (x1root)
      [black,below=0.15cm] {$\small \bx^1$}; 
      \fill[black]  (2.5, 0.5) circle (2pt) node (x4root)
      [black,below=0.15cm] {$\small \bx^4$}; 
      \fill[black]  (2, 0.5) circle (2pt) node (x3root)
      [black,below=0.15cm] {$\small \bx^3$};

      \draw[snake=brace, mirror snake, color = darkgreen, segment
      amplitude=3pt, line width=0.5pt] (x1root.south west) --
      (x4root.south east);
      
      \draw[-, darkgreen, thick] (5, 1) -- (6, 1);
      \draw[-, darkgreen, thick] (5, 1) -- (5, 0.5);
      \draw[-, darkgreen, thick] (5.5, 1) -- (5.5, 0.5);
      \draw[-, darkgreen, thick] (6, 1) -- (6, 0.5);

      \fill[black] (5, 0.5) circle (2pt) node (x1g1)
      [black,below=0.15cm] {$\small \bx^1$}; 
      \fill[black]  (5.5, 0.5) circle (2pt) node (x3g1)
      [black,below=0.15cm] {$\small \bx^3$}; 
      \fill[black]  (6, 0.5) circle (2pt) node (x4g1)
      [black,below=0.15cm] {$\small \bx^4$}; 

      \draw[-, orange, thick] (3.5, 2.75) -- (4, 2.75);
      \draw[-, orange, thick] (3.5, 2.75) -- (3.5, 0.5);
      \draw[-, orange, thick] (4, 2.75) -- (4, 0.5);

      \fill[black] (3.5, 0.5) circle (2pt) node (x2g2)
      [black,below=0.15cm] {$\small \bx^2$}; 
      \fill[black]  (4, 0.5) circle (2pt) node (x5g2)
      [black,below=0.15cm] {$\small \bx^5$}; 

    \end{tikzpicture}
  }
  \subfigure[Compressed representation]{
    \label{fig:p-h:compressed}
    \begin{tikzpicture}
      \draw[color = bgfig, draw=gray] (0, -0.5) rectangle
      (4, 3.75);

      \fill[blue] (0.5, 0.5) circle (2pt) node (xt1)
      [black,below=0.15cm] {$\small \textcolor{blue}{\rho_1\tilde{\bx}^1}$};   
      \fill[orange]  (2, 0.5) circle (2pt) node (xt2)
      [black,below=0.15cm] {$\small
        \textcolor{orange}{\rho_2\tilde{\bx}^2}$};  
      \fill[darkgreen]  (3.5, 0.5) circle (2pt) node (xt3)
      [black,below=0.15cm] {$\small
        \textcolor{darkgreen}{\rho_3\tilde{\bx}^3}$};  
      
      \draw[-, black, thick] (2, 3) -- (2, 3.5);

       \draw[-, black, thick] (0.5, 3) -- (3.5, 3);

       \draw[-, blue, thick] (0.5, 0.5) -- (0.5, 3);

       \draw[-, orange, thick] (2, 0.5) -- (2, 3);

       \draw[-, darkgreen, thick] (3.5, 0.5) -- (3.5, 3);
    \end{tikzpicture}
  }
  \caption{Dimension reduction strategy. (a) Original hierarchical
    tree with an example for 5 variables. (b) Expanded representation
    of the tree with all possible weighted groups derived from the
    original hierarchy. The group in blue gathers the variables
    contained in the groups in orange and green. (c) Compressed
    representation of the tree after construction of the
    supervariables.}
  \label{fig:process-hierarchy}
\end{figure*}

\subsection{Using the structure efficiently}

Different approaches for finding an optimal number of clusters may be
envisaged when looking for the optimal cut in a tree structure
obtained by hierarchical clustering (see for instance
\citep{Milligan:Ps1985} or \citep{Gordon:1999}). Whatever the
approach, finding this optimal cut necessarily involves a systematic
exploration of different levels of the hierarchy.
Our alternative strategy for bypassing this expensive exploration is
as follows:

\begin{enumerate}[(a)~]
\item \label{step:a}  Expanding the hierarchy, considering all possible
  groups at a single level;
\item \label{step:b} Assigning a weight to each group based on the
  distances between two consecutive groups in the hierarchy;
\item \label{step:c} Compressing each group into a supervariable.
\end{enumerate}

The different steps in this strategy are illustrated in
Figure~\ref{fig:process-hierarchy}, from the original tree structure
in Figure~\ref{fig:p-h:original} to the final flattened, weighted,
compressed representation shown in Figure~\ref{fig:p-h:compressed}.

\subsubsection*{Expanding the hierarchy \hfill{(\ref{step:a})}}

To reduce the dimension of
Problem~\eqref{eq:compact_vector_model}, the first step consists in
flattening the respective tree structures obtained on views $\vu$ and
$\vd$ so that only one group structure remains. Each group of
variables defined at the deepest level
may thus be included in other groups of larger scales, as shown in
Figure~\ref{fig:p-h:expanded}.

\subsubsection*{Assigning weights to the groups \hfill{(\ref{step:b})}}

To keep track of the tree structure, an additional measure may be
included to quantify
the loss of information between two successive levels.  More
specifically, for a tree structure of height $H$ and for
$1 \leq h \leq H-1$, we define $s_h$ as the gap between heights $h$
and $h-1$. Using a similar methodology to \cite{Grimonprez:2016} for
the multi-layer group Lasso, we define this quantity as
$\displaystyle{\rho_h = 1 / \sqrt{s_h}}$. The process is shown in
Figure~\ref{fig:p-h:original} and~\ref{fig:p-h:expanded}.

\subsubsection*{Compressing the data \hfill{(\ref{step:c})}}

To summarize each group of variables the mean, the median, or other
quantiles may be used, as well as more sophisticated representations
based on eigenvalue decomposition, such as the first factor of a
Principal Component Analysis.

\subsection{Identification of relevant supervariables}

With the aid of this compressed representation we can uncover relevant
interactions using a multiple testing strategy.

\subsubsection*{Selection of supervariables}

Compression is a key ingredient in reducing significantly the
dimension of Problem~\eqref{eq:compact_vector_model}. We take this
a step further with an additional feature selection
process applied to the compressed variables, as described at the beginning
of this section, in order to preprocess the data using screening rules and/or
applying a Lasso optimization on each view $\vu$ and $\vd$:
\begin{align*}
  \argmin_{\bbeta_{\vu}}  & \;
                            \sum_{i=1}^n \left(y_i - \tbx_{i}^{\vu}
                            \bbeta_{\vu} \right)^2   \  + \
                            \lambda_\vu \sum_{\ivu=1}^{N_{\vu}} \rho_\ivu
                            |\beta_{\ivu}| \,,
\end{align*}
and
\begin{align*}
  \argmin_{\bbeta_{\vd}} & \;
                           \sum_{i=1}^n \left(y_i - \tbx_{i}^{\vd}
                           \bbeta_{\vd} \right)^2  \ + \
                           \lambda_\vd \sum_{\ivd=1}^{N_{\vd}} \rho_\ivd
                           |\beta_{\ivd}| \,,
\end{align*}
with penalty factors defined by
$\displaystyle{\rho_\ivu = 1 / \sqrt{s_\ivu}}$ and
$\displaystyle{\rho_\ivd = 1 / \sqrt{s_\ivd}}$, as explained in
Section~\ref{sec:structure}.

\ifstabs %
{This step for selecting the supervariables in the two complementary
  datasets can be subject to instability when setting the amount of
  selection. The method can be improved further in terms of model
  consistency by using resampling techniques \citep{Bach:ICML08,
    MB:JRSSB10, Hofner:StabSel:2015}. This has been implemented in
  SICOMORE with the \texttt{R} package
  {stabs}~\citep{Hofner:pkg_stabs:2017}}.  %
\fi

\subsubsection*{Linear model testing}

For the purpose of feature selection the relevant interactions may be
uncovered separately by considering each selected group $\ivu \in \gvu$
coupled with each selected group $\ivd \in \gvd$ in a linear model of
interaction and by performing a hypothesis test (a standard $t$-test
for instance) on each parameter $ {\theta_{\ivu\ivd}}$:
\begin{equation}
  \label{eq:compact_single_interaction_model}
  y_i = \tilde{x}_i^\ivu \beta_{\ivu}  +
  \tilde{x}_i^\ivd \beta_{\ivd}  +
  \left(\tilde{x}_i^\ivu  \cdot
    \tilde{x}_i^\ivd\right)
  {\theta_{\ivu\ivd}} + \, \epsilon_i \,.
\end{equation}

This strategy has the advantage of highlighting all the potential
interactions between the selected simple effects in an exploratory
rather than a predictive analysis perspective.
It can also be seen as an alternative way of shortcutting
Problem~\eqref{eq:compact_vector_model}, in that it involves $N_I$
problems of dimension $3$ rather than a potentially large problem of
dimension $N_{\vu} + N_{\vd} + N_I$.
Finally, by construction, this selection scheme preserves strong
dependencies.


\section{Numerical simulations}
\label{sec:XP_simu}

{We present some numerical simulations to assess SICOMORE's ability to
  uncover relevant interactions. We compare our approach with two
  other methods, namely \textbf{MLGL}~\citep{Grimonprez:2016} and
  \textbf{glinternet}~\citep{lim2015learning}. These two methods will
  be described in more detail later in the section. Both are available
  as \texttt{R} packages on the CRAN
  platform~\citep{Grimonprez:pkg_MLGL:2020, Lim:pkg_glinternet:2019}.}

{These numerical simulations are designed to study several aspects of
  SICOMORE:
  \begin{itemize}
  \item The ability to recover relevant interactions will be observed
    on different configurations with respect to the sample sizes, the
    noise, and the number of true interactions.
  \item The impact of the weighting scheme will be shown with two
    versions of our approach, using both weighted and unweighted
    supervariables.
  \item The impact of the compression scheme will be compared to MLGL
    using the same structure but with the initial variables.
  \item Finally, a dedicated simulation sketches the running times
    necessary for each method to reach convergence when the dimension
    of one of the matrices grows. To allow the comparison of SICOMORE
    with MLGL or glinternet, the dimensions of the simulated
    matrices have been kept between a few hundred and a few thousand.
  \end{itemize}
}

\subsection{Data generation}

\subsubsection*{Generation of metagenomic and genomic data matrices}

\begin{figure}
 \centering%
 \subfigure[Correlation matrix of $\bX^\vu$]{%
   \fcolorbox{gray}{white}{\includegraphics[scale=.2]{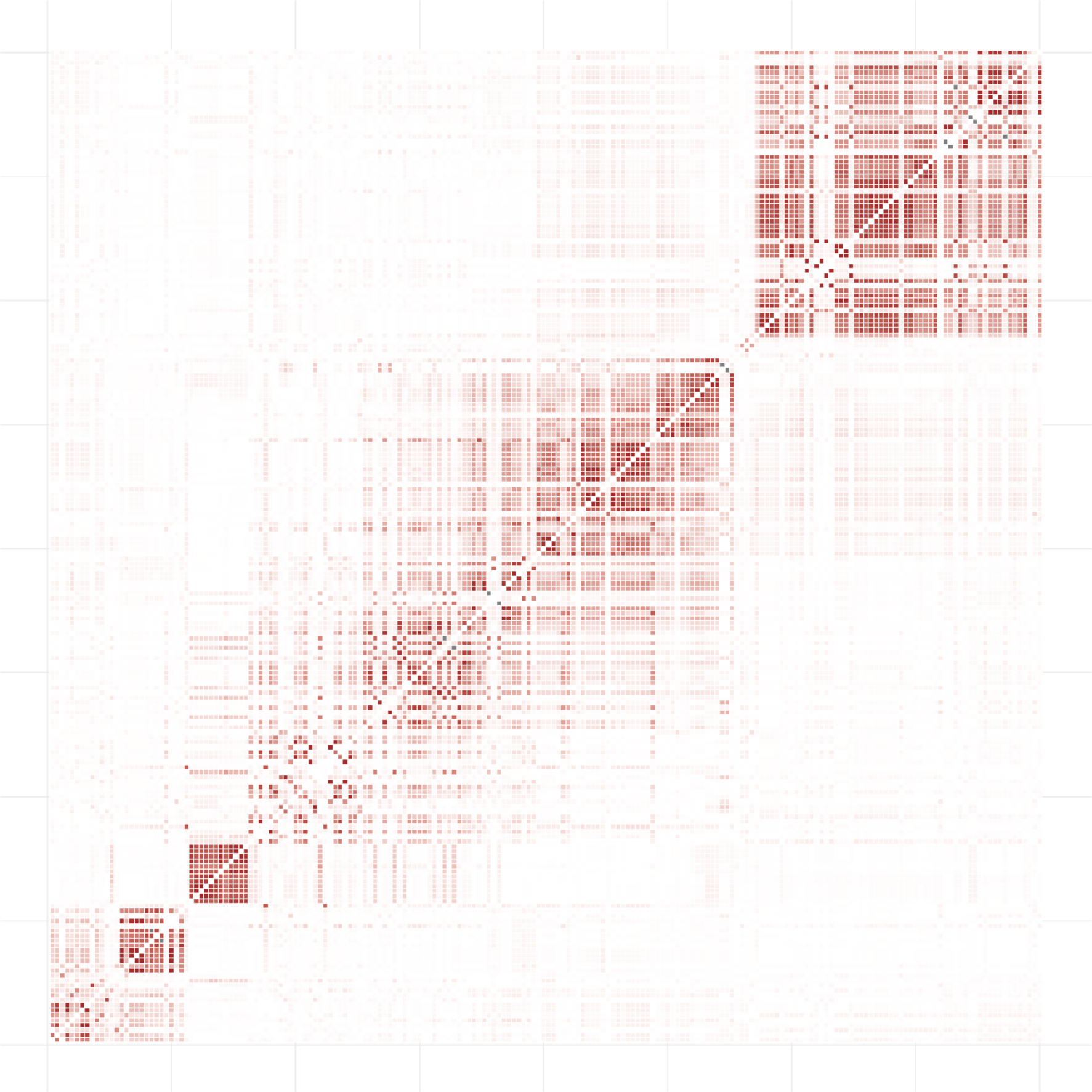}}
   \label{fig:XG}
 }%
 \subfigure[Correlation matrix of $\bX^\vd$]{%
   \fcolorbox{gray}{white}{\includegraphics[scale=.2]{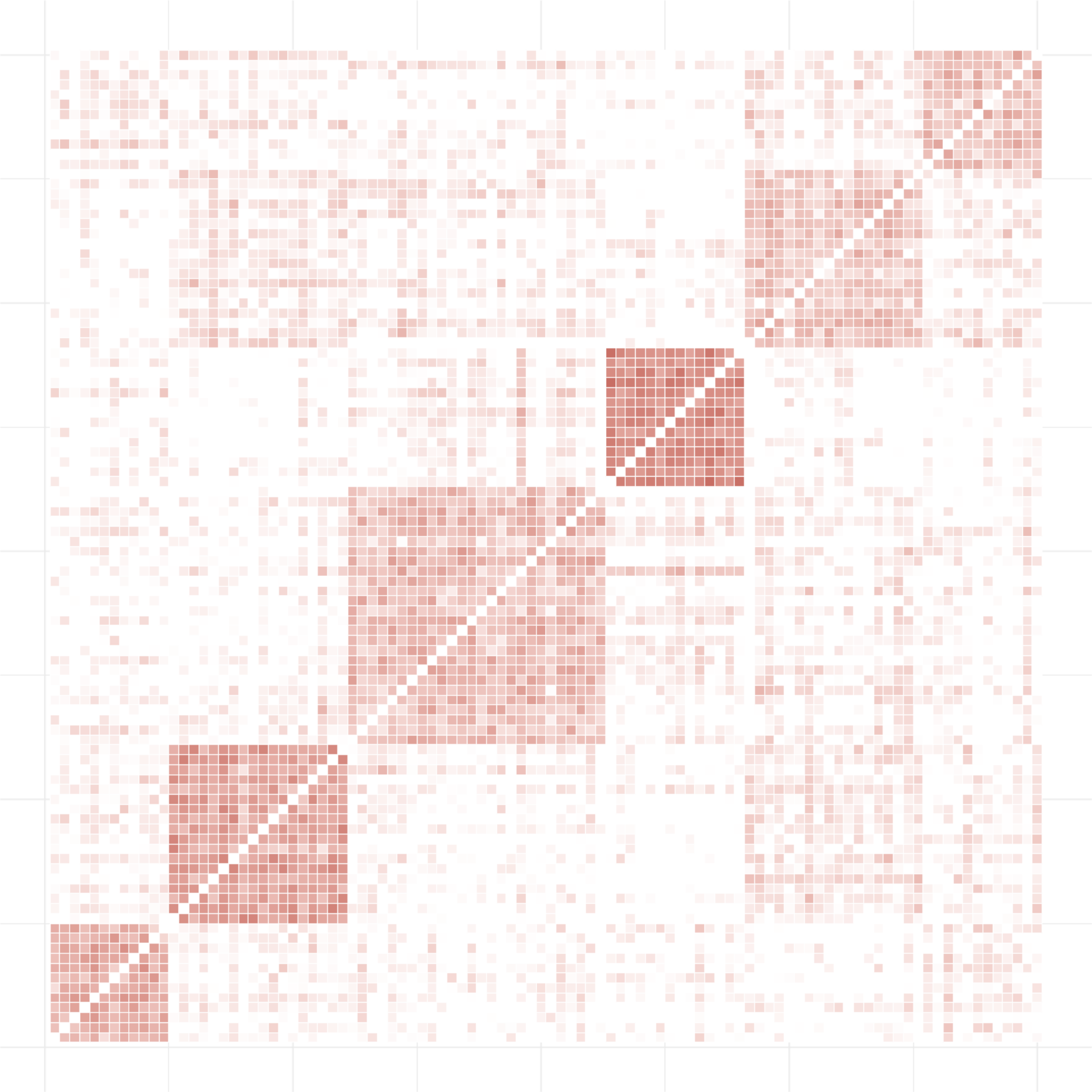}}
   \label{fig:XM}
 }
 \caption{Examples of group structures: correlations observed on (a)
   genomic data $\bX^\vu$ and (b) metagenomic data $\bX^\vd$.}
 \label{fig:simureal_data}
\end{figure}

\paragraph{Genomic data.} To obtain a matrix $\bX^{\vu}$ resembling
real genomic data we used \texttt{HAPGEN2} software~\citep{SU:BIO2011,
  Su:pkg_HAPGEN:2011}, which can simulate an entire chromosome
conditionally on a reference set of population haplotypes (from
HapMap3) and an estimate of the fine-scale recombination rate across
the region, so that the simulated data share similar patterns with the
reference data. We generated chromosome 1 using the haplotype
structure of CEU population (Utah residents with Northern and Western
European ancestry from the CEPH\footnote{ \url{http://www.cephb.fr}.}
collection) as the reference set, and we selected $D_\vu=200$
variables from this matrix to obtain the simulated dataset. An example
of the linkage disequilibrium structure among the simulated SNPs is
shown in Figure~\ref{fig:XG}.

\paragraph{Metagenomic data.}
The data matrix $\bX^{\vd}$, with $D_\vd=100$ variables, was generated
using a multivariate Poisson log-normal distribution
\citep{aitchison1989multivariate} with group structure dependencies.
The Poisson log-normal model is a latent Gaussian model where latent
vectors $ \vZ_i \in \IR^{D_\vd} $ are drawn from a multivariate normal
distribution
$$\vZ_i \sim \mathcal{N}_{D_\vd}(0, \bSigma) \ , $$
and where $\bSigma$ is a covariance matrix that can give a
correlation structure between the variables.  The {random variable
  $\vX^{\vd}_i$} related to the centered phenotypic count data is then
drawn from a Poisson distribution conditionally on $\vZ_i$
$$\vX^{\vd}_{ij}|\vZ_{ij} \sim \mathcal{P}\left( e^{\mu_j +
    \vZ_{ij}}\right).$$

The group structure shown in Figure~\ref{fig:XM} was obtained by
drawing a latent multivariate normal vector using a covariance matrix
such that the correlation level between the latent variables in a
group are between 0.5 and 0.95.  Simulating in this way gives a matrix
of count data with a covariance structure close to what is observed
with metagenomic data. As described in Section \ref{sec:preprocess},
we computed the proportions for each of the random variables and
transformed them using centered log-ratios.

\setlength{\fboxrule}{0.25pt}

\subsubsection*{Generation of the phenotype}
For all simulations we used a fixed value of $N_{\vd} = 6$ groups for
the matrix $\bX^{\vd}$. For the matrix $\bX^{\vu}$, since
\texttt{HAPGEN2} does not allow the group structure to be controlled
exactly, we used the gap statistic \citep{TGH:JRSS2016} to identify a
number of groups in the hierarchy. For instance, in
Figure~\ref{fig:XG}, the gap statistic identified $N_{\vu} = 16$
groups. The supervariables were then calculated using averaged groups
of variables to obtain the two matrices of supervariables,
$\tilde{\bX}^{\vu}$ and $\tilde{\bX}^{\vd}$.

To generate the phenotype, we considered a data structure for which
the data to regress was generated using supervariables according
a linear model with interactions of the form:

\begin{equation}
  \label{eq:phenotype_gen}
  y_i =
  \sum_{\ivu \in \mathcal{S}^{\vu}}\tilde{x}_i^\ivu \beta_{\ivu} +
  \sum_{\ivd \in \mathcal{S}^{\vd}} \tilde{x}_i^\ivd \beta_{\ivd} +
  \sum_{\ivu \in \mathcal{S}^{\vu}}\sum_{\ivd \in \mathcal{S}^{\vd}}
  \underbrace{\left(\tilde{x}_i^\ivu  \cdot
      \tilde{x}_i^\ivd\right)}_{\phi^{\ivu\ivd}_i}
  {\theta_{\ivu\ivd}} + \, \epsilon_i \,,
\end{equation}
where $\mathcal{S}^{\vu}$ and $\mathcal{S}^{\vd}$ are subsets of
randomly chosen effects from the matrices $\tilde{\bX}^{\vu}$ and
$\tilde{\bX}^{\vd}$ respectively, $\tilde{x}_i^\ivu $ is the $i^{th}$
sample of the $\ivu$ effect and $\beta_{\ivu} $ its corresponding
coefficient, and $\tilde{x}_i^\ivd $ is the $i^{th}$ sample of the $\ivd$
effect and $\beta_{\ivd} $ its corresponding coefficient.  Finally,
$\theta_{\ivu\ivd}$ is the interaction between variables
$\tilde{x}_i^\ivu$ and $\tilde{x}_i^\ivd$.

We considered $I \in \{1,3,5,7,10\}$ true interactions between some
supervariables to generate the phenotype such that $I$ 
blocks of the coefficients of $\theta_{\ivu\ivd}$ have non zero
values.
The process was repeated 30 times for each couple of parameters in
$N=\{50, 100, 200\} \times sd(\bepsilon)=\{0.5, 1, 2\}$.


\subsection{Comparison of methods}
\label{sec:comparison}

{In accordance with the outline given in the preamble of
  Section~\ref{sec:XP_simu}, we were seeking to assess the ability of
  SICOMORE, in comparison with {MLGL} and {glinternet}, to uncover
  true causal interactions. For this purpose, we needed to reshape the
  datasets provided to the two methods as we now describe below.}

It is worth mentioning that SICOMORE is an approach that draws on the
work of~\cite{PHT:Biostat07} and {MLGL}~\citep{Grimonprez:2016}, with
an explicit design for detecting interactions. We explore two settings
: $\rho$-SICOMORE and SICOMORE, which correspond respectively to the
method described in section~\ref{sec:implementation} using
$\displaystyle{\rho_h = 1 / \sqrt{s_h}}$ and $\rho_h = 1$,
$\forall h$.

\subsubsection*{Multi-Layer Group Lasso \hfill (MLGL)}

\cite{Grimonprez:2016} defines {MLGL} as a two-step procedure that
combines a hierarchical clustering with a group Lasso regression. It
is a weighted version of the overlapping group Lasso
\citep{JOV:ICML09} which 
performs variable selection on multiple group partitions defined by
the hierarchical clustering. A weight is attributed to each possible
group identified at all levels of the hierarchy, as described in
Section~\ref{sec:implementation}(\ref{step:b}).  This weighting scheme
favors the creation of groups associated with large gaps in the
hierarchy.

The model of interactions is fitted with weights on the groups defined
by the expanded representation of the two hierarchies {using the
  initial variables}, as illustrated in Figure \ref{fig:p-h:expanded}.
The ability of MLGL to uncover real interactions is evaluated
positively if it selects the correct interaction terms between two
groups of variables at the right level in both hierarchies.

{It should be noted that here {MLGL} is not being evaluated in a
  context for which it was intended, since {MLGL} examines the
  different levels of a hierarchical structure using \emph{all}
  variables. This approach is not well suited in a high-dimensional
  setting and still less in a model of interactions. But, as we
  explained at the beginning of Section~\ref{sec:XP_simu}, this
  comparison with MLGL is intended to shed light on the impact of the
  compression applied to the variables in SICOMORE.}

\subsubsection*{Group Lasso interaction network \hfill (glinternet)}

\cite{lim2015learning} introduced {glinternet}, a procedure that
considers pairwise interactions in a linear model in a way that
satisfies strong dependencies between main and interaction effects:
whenever an interaction is estimated to be non-zero, its two
corresponding main effects are also included in the
model.

It fits a hierarchical group Lasso model, with constraints on the main
and interactions effects, as specified in Section
\ref{sec:recover-interactions}, and it accommodates the strong
dependency hypothesis by adding an appropriate penalty to the loss
function (we refer the reader to \citep{lim2015learning} for more
details on the form of the penalty). For very large problems (with a
number of variables $\geq 10^5$), the group Lasso procedure is
preceded by a screening step that gives a candidate set of main
effects and interactions.

Since this method can only work at the level of variables, we needed
to include a group structure into the analysis, and so we decided to
fit the glinternet model on the compressed variables and to constrain
the model to only fit the interaction terms between the supervariables
of the two matrices $\tilde{\bX}^{\vu}$ and $\tilde{\bX}^{\vd}$. We
explicitly removed all interaction terms between supervariables
belonging to the same data matrix.

To ensure that our comparison of SICOMORE was fair, we considered two
options, namely {GLtree} and {GLgap}. The GLtree option works on the
unweighted compressed representations of the two hierarchies
(Figure~\ref{fig:p-h:compressed}) and thus takes into account all the
possible interactions between the supervariables of the two
datasets. In contrast, the GLgap option considers only the
interactions between the compressed variables constructed at a
specific level in the hierarchies, chosen by the gap statistic.  Given
that $D^{\vu}$ and $D^{\vd}$ are the numbers of variables in
$\bX^{\vu}$ and $\bX^{\vd}$, the dimension of the matrices
$\tilde{\bX}^{\vu}$ and $\tilde{\bX}^{\vd}$ in GLtree are respectively
$\tilde{D}^{\vu} = D^{\vu} + (D^{\vu}-1)$ and
$\tilde{D}^{\vd} = D^{\vd} + (D^{\vd}-1)$.~\footnote{{In GLtree, a
    matrix $\tilde{\bX}$ is created using the initial $D$ variables,
    and the $(D-1)$ groups of variables of the dendogram from the
    hierarchical clustering are added as compressed
    features.}} 
Consequently, for GLtree the number of interactions to be examined is
$\tilde{D}^{\vu} \times \tilde{D}^{\vd}$, while for GLgap this number
will depend on the level chosen by the gap statistic, but it will
necessarily be smaller since this option considers only a specific
level of the hierarchy. In the numerical simulations, given that
$D^{\vu} = 200$ and $D^{\vd}=100$, the use of strong rules to discard
variables is therefore not necessary.

\subsection{Evaluation metrics}

\begin{figure}[thb]
  \centering
  \fbox{\includegraphics[width=0.7\textwidth]{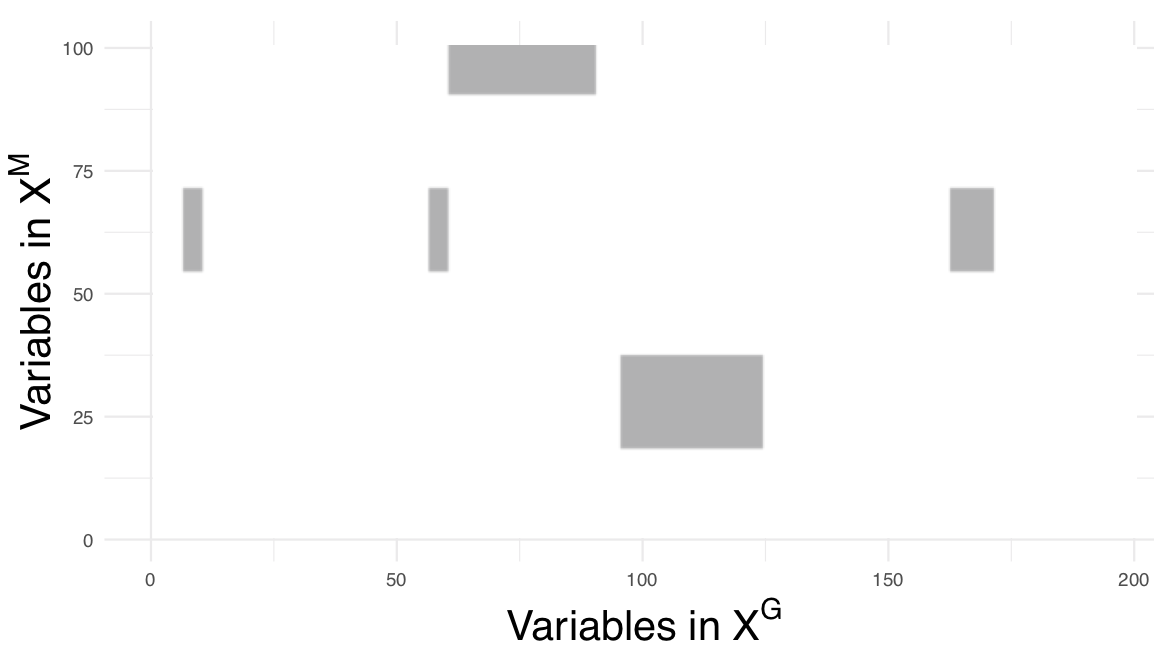}}
  \caption{\textbf{Illustration of the true interaction matrix} $\btheta$
   with $I = 5$, $\sigma = 0.5$ and $n=100$. Each non-zero value in this
   matrix is considered as a true interaction between two variables.}
   \label{fig:ggheat_theta}
 \end{figure}

\begin{figure}[thb]
  \centering
  \fbox{\includegraphics[width=0.95\textwidth]{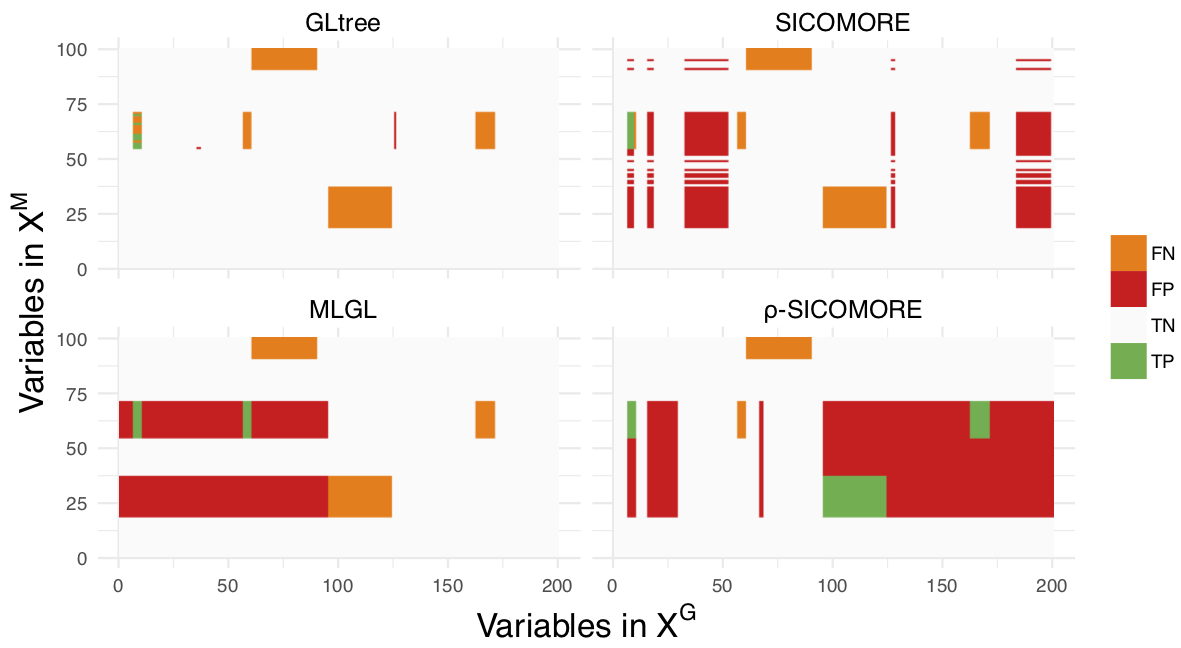}}

  \caption{\textbf{Confusion matrices of interactions}
    $\hat{\theta}_{jj'}$ for the different methods, using the
    following simulation parameters: $I = 5$, $\sigma = 0.5$,
    $n=100$. We can see from this example that MLGL and $\rho$-SICOMORE
    behave similarly, with very large genomic regions identified.
    SICOMORE tends to work with smaller genomic and metagenomic
    regions.}
   \label{fig:theta2by2}
\end{figure}

For each run we evaluated the quality of the variable selection using
Precision and Recall.  More precisely, we compared the true
interaction matrix ${\btheta}$ that we used to generate the phenotype
with the estimated interaction matrix $\hat{\btheta}$ computed for each
model.

For all possible $D_{\vu} \times D_{\vd}$ interactions, with $\theta_{jj'}$
the interaction term between variable $j \in \bX^{\vu}$ and
variable $j' \in \bX^{\vd}$, we determined the following confusion
matrix:
\begin{center}
  \begin{tabular}{c@{\hspace{-.25cm}} | c | c@{\hspace{-.025cm}}}

    & $\hat{\theta}_{jj'}=0$ &
                                    $\hat{\theta}_{jj'}\neq0$ \phantom{\Huge p} \\
    \hline
    ${\theta}_{jj'}=0$ \phantom{\Huge p}
    & True Negative & False Positive \\
    \hline
    ${\theta}_{jj'}\neq0$ \phantom{\Huge p}
    & False Negative & True Positive \\

  \end{tabular}
\end{center}

\bigskip The performances are measured with
$\text{Precision}={\frac{TP}{FP+TP}}$ and
$\text{Recall}=\frac{TP}{FN+TP}$.  An example of the interaction
matrix $\hat{\btheta}$ is shown in Figure \ref{fig:ggheat_theta} for
$I=5$ blocks in interaction.

 Here, a \emph{true positive} corresponds to a significant $p$-value
 on a true causal interaction, a \emph{false positive} to a
 significant $p$-value on a noise interaction, and a \emph{false
   negative} to a non-significant $p$-value on a true causal
 interaction.

 For the three tested methods we corrected for multiple testing by
 controlling the family-wise error rate with the Holm-Bonferroni
 method. Even though it is known to be stringent, we chose the
 Holm-Bonferroni method to adjust for multiple testing because the
 number of hypothesis tests that needed to be performed for our
 simulation was quite low.  In a high-dimensional context, for example
 in analyzing real microarray 
 data, the Benjamini-Hochberg method would be preferable for
 controlling the false discovery rate.

\subsection{Performance results}

\begin{figure}[htb]
  \centering
  \subfigure[]{
    \label{fig:Precision}
    \includegraphics[width=0.92\textwidth]{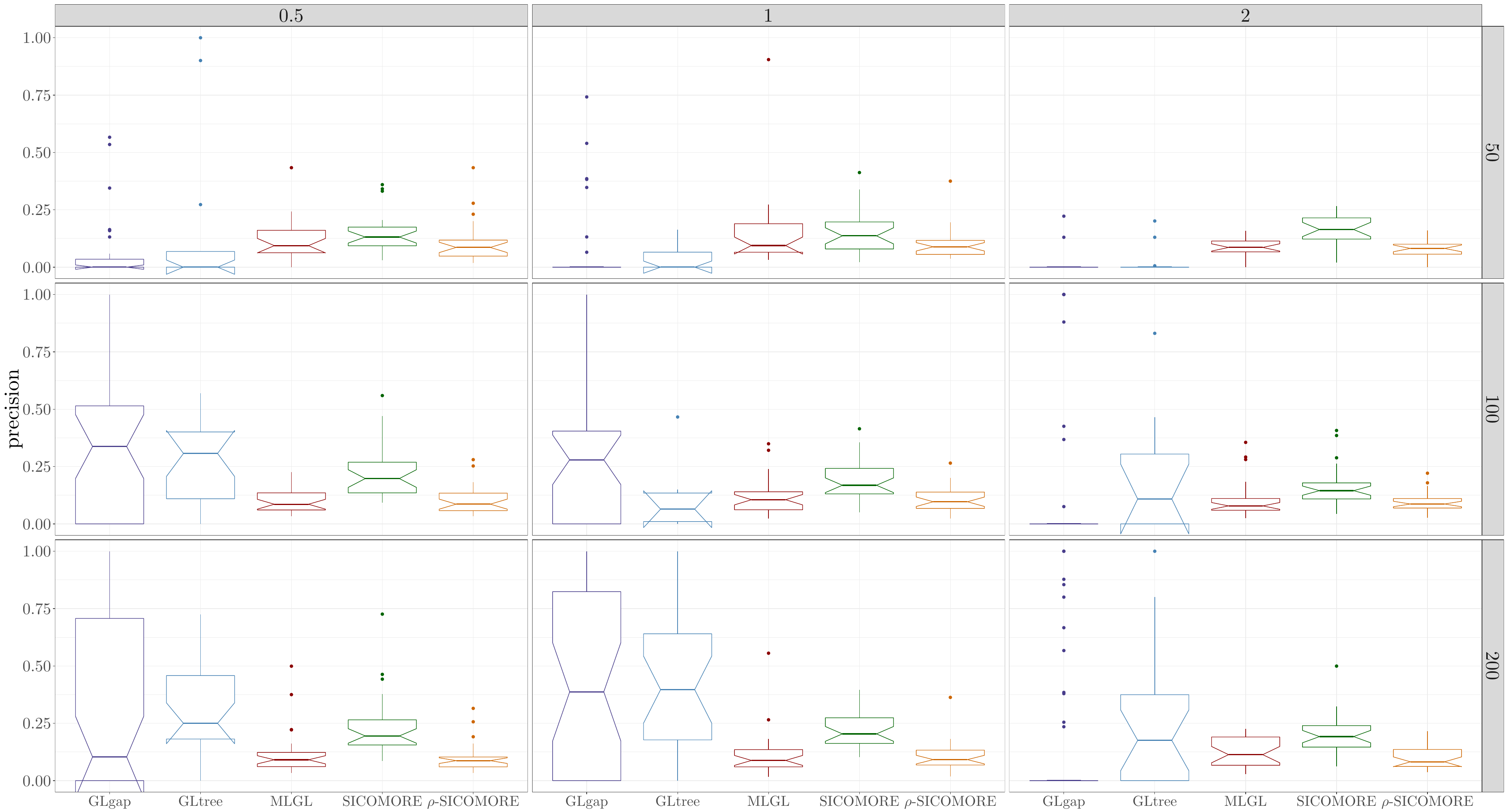}
  }\\
  \subfigure[]{
    \label{fig:Recall}
    \includegraphics[width=0.92\textwidth]{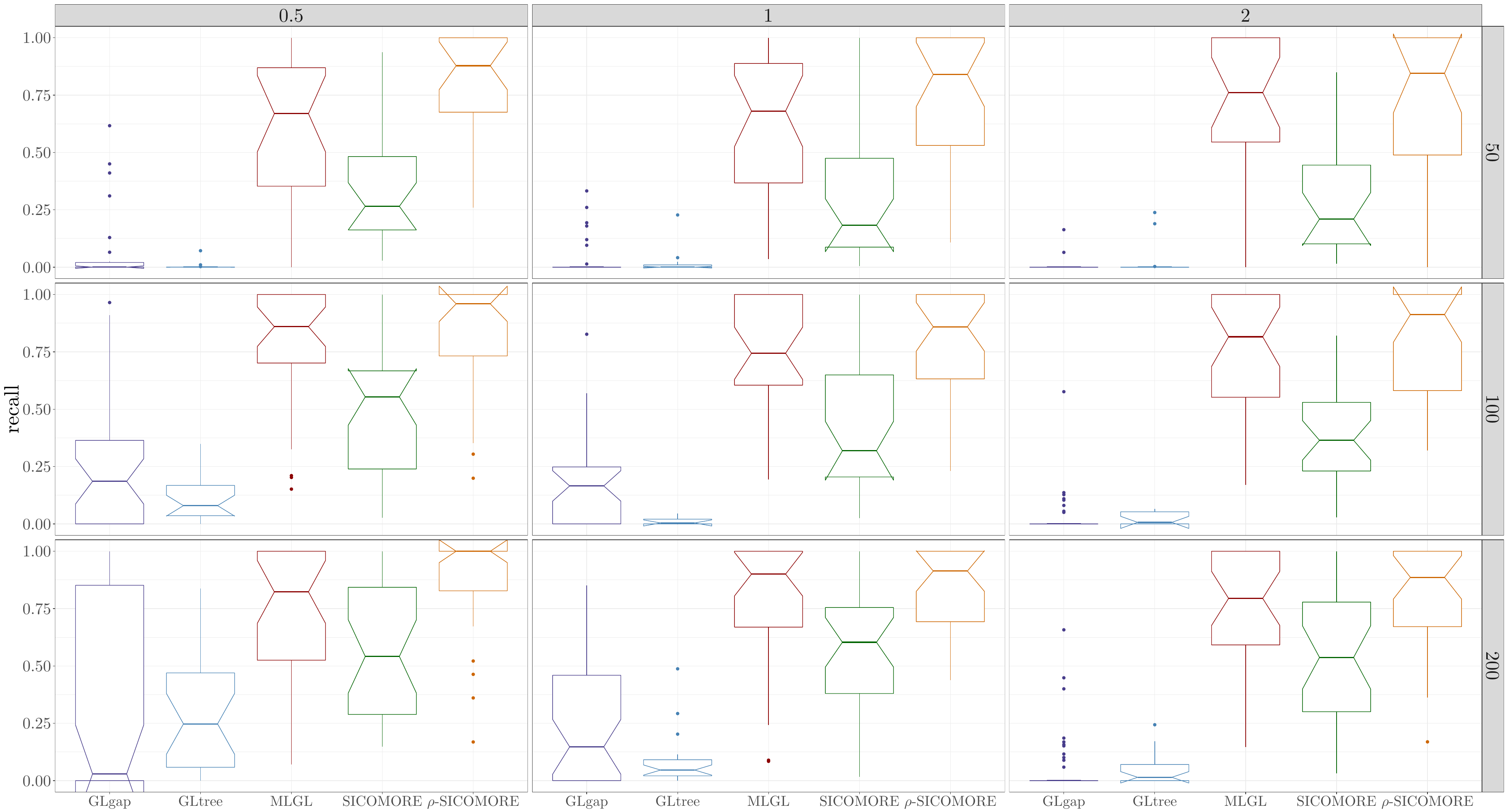}
  }
  \caption{Boxplots of (a) Precision and (b) Recall results obtained on the
    numerical simulations with a Bonferroni-Holm correction for $I=7$
    blocs. The lines correspond to different numbers of
    observations (top: $N=50$, middle: $N=100$ and bottom: $N=200$),
    and the columns correspond to levels of difficulty of the problem (left:
    $\epsilon=0.5$, middle: $\epsilon=1$ and right: $\epsilon=2$). The
    boxplots are best seen in colors: from the left to the right,
    GLgap is in purple, GLtree is in blue, MLGL is in red, SICOMORE is
    in green, $\rho$-SICOMORE is in orange.}
  \label{fig:precision_recall}
\end{figure}

The performances of the different methods in uncovering true causal
interactions are shown in figures~\ref{fig:Precision} (for Precision)
and \ref{fig:Recall} (for Recall). For the sake of clarity we show
only the results for $I=7$ blocks of variables in interaction. The
results for $I \in \{1,3,5,10\}$ are provided in
Appendix~\ref{sec:supp} as supplementary results. The plots in Figure
\ref{fig:theta2by2} represent the uncovered confusion matrices of
interaction $\theta_{gm}$ corresponding to one particular set of
simulation parameters ($I = 5$, $\sigma = 0.5$, $n=100$) for each of
the compared methods.

The Recall results show that MLGL and $\rho$-SICOMORE are good at
uncovering true positive interactions, with $\rho$-SICOMORE performing
better overall.  SICOMORE performs less well because it favours the
selection of small groups that are only partly contained in the groups
that generate the interactions. This indicates that MLGL and
$\rho$-SICOMORE have an effective weighting scheme.  GLgap is unable
to uncover relevant interactions, but here the way the structure
between variables is defined using the gap statistic differs from the
other methods.  The Precision results show that all methods perform
poorly, with a significant number of false positive interactions. MLGL
and $\rho$-SICOMORE tend to select groups of variables and
supervariables that are too high in the tree structure, giving rise to
false positives that are spatially close to the true
interactions. SICOMORE, which, as explained above, favours small
groups, gives fewer false positives of this kind. The behaviour of
GLgap may vary according to the selected cut with the gap statistic
into the tree structure, while the GLtree option has slightly better
precision. Note that this improved precision may be the consequence of
the additional information provided from our group definition.  The
glinternet method is mostly unable to uncover the true interactions
correctly, whether the compressed or the original representation is
used.

\subsection{Computation time}

In order to reduce the computation time required to run our algorithm,
we chose to restrict the search space.
It is limited to the area of the tree where the jumps in the hierarchy
are the largest, and the number of groups to be evaluated is
arbitrarily set to five times the number of initial features. This
reduces the number of variables to be fitted in the Lasso regression
but does not affect performance regarding Recall and Precision.

We compared the computational performance of our method with the two
others by varying the number of variables in $\tilde{\bX}^{\vu}$. We
repeated the number of evaluation five times for each size of
$\tilde{\bX}^{\vu}$ and averaged the computation time.


\begin{table}[htbp]
\centering
\begin{tabular}{r| rrrrrrrr}
  \hline
    $N_{\vu}$ &
                      50 & 100 & 500 & 1000 & 1500 &
                                                     2000 & 3000 & 4000 \\
  \hline
  $\rho$-SICOMORE & 0.01 & 0.01 & 0.02 & 0.03 & 0.03 & 0.04 & 0.05 &
                                                                     0.06 \\
  SICOMORE & 0.21 & 0.34 & 0.82 & 0.76 & 0.75 & 0.96 & 0.93 & 1.09 \\
  MLGL & 0.06 & 0.09 & 3.35 & 0.86 & 3.12 & 4.52 & 8.02 & 24.20 \\
  GLtree & 0.07 & 0.28 & 0.67 & 3.83 & 11.69 & 26.31 & 88.17 &
                                                                   210.64\\
  \hline
\end{tabular}
\medskip
\caption{Average computation time (in minutes) over 5 replicates for
  varying dimensions of $\tilde{\bX}^{\vu}$, with the dimension of
  $\tilde{\bX}^{\vu}$ being fixed ($N_{\vd}=6$).} 
\label{tab:time}
\end{table}

We can conclude from the results presented in Table \ref{tab:time}
that two methods, glinternet and MLGL, are unsuitable for large-scale
analyses of genomic data, since computation time starts to rise
steeply once the number of variables exceeds a few thousand.  The
computation time of $\rho$-SICOMORE and SICOMORE is drastically
reduced compared to MLGL or glinternet, with $\rho$-SICOMORE having a
slight advantage due to the weighting scheme that induces faster
elimination of non relevant supervariables.


\section{Application on the {rhizosphere bacterial communities} of
  \textit{Medicago truncatula}}
\label{sec:XP_INRA}

For an implementation of our algorithm on real data we chose to study
the interactions between the genome of \textit{Medicago truncatula}
and the metagenome {(16S rRNA gene sequencing) of its rhizosphere
  bacterial community}.  {We were seeking to identify significant
  interactions in order to better understand the effect of both the
  plant genome and the rhizosphere bacterial microbial community on
  plant growth.}

For this purpose, a core collection of {155} accessions (all from
INRAE-Montpellier) were grown in a controlled environment and
phenotyped for several traits related to the plant growth and
nutritional strategy:
\begin{itemize}
\item Total {Dry} Biomass (TDB).
\item Root Total {Dry} Biomass Ratio (RTDBR).
\item Specific Nitrogen Uptake (SNU) expressed as $mg$ of $N.g^{-1}$
  of belowground biomass per day.
\end{itemize}

{In addition to the phenotypic measurement, the rhizosphere of each
  accession was also analyzed to determine the bacterial diversity and
  composition ({see Appendix~\ref{sec:supp_data}}).}
{The metabarcoding raw data is available in the European Nucleotide
  Archive (ENA) EMBL-EBI database system under project accession
  PRJEB25849.}

A total of {15617} different {bacterial} OTUs were found in the
rhizosphere of the plants.
{The different OTUs were pooled according to their taxonomic
  affiliation at the genus level, and a total of 329 genera were thus
  analyzed.}

The {155} sequenced accessions, extracted from
\url{http://www.medicagohapmap.org}, were genotyped with a DNA
microarray chip, giving a total of 6 372 968 SNPs after 3\% MAF,
multiallele SNP exclusion and minimum count (100) filtering. The
missing values were imputed using the \texttt{snp.imputation} function
from the \texttt{R} package
snpStats~\citep{Clayton:pkg_snpStats:2019}.  Given two sets of SNPs
typed in the same subjects, this function computes rules that can be
used to impute one set from the other in a subsequent sample. By
discarding any SNP that had too many missing values to be completely
imputed, we reduced the size of the data to 2 148 505 SNPs.

The positions of SNPs inside or in the vicinity of genes ($\pm$ 2Kb) were
extracted from context files downloaded from
\url{http://www.medicagohapmap.org}. A Singular Enrichment Analysis
was conducted using an exact Fisher test with the \texttt{R} package
{topGO}~\citep{Alexa:pkg_topGO:2019} and GO term annotation from
\url{http://www.medicagogenome.org}.

The algorithm requires several hyper-parameters to be chosen in
order to run properly:
\begin{itemize}
\item \textbf{Aggregating function}: %
  For the genomic and the metagenomic data, we defined the {mean}
  value of the group as supervariable.%

\item \textbf{Clustering algorithm}: For the metagenomic data we used
  a hierarchical clustering using Ward’s distance as the measure of
  similarity. For the genomic data we used a spatially constrained
  hierarchical clustering algorithm that integrates the linkage
  disequilibrium as the measure of dissimilarity.
\ifstabs
\item {\textbf{Stability selection}: The parameters of the function
    stabs in SICOMORE for the metagenomic data were fixed to
    $\mathtt{B}=300$ subsampling replicates, with the frequency of
    selection of the supervariables on the replicates
    $\mathtt{cutoff}=0.7$. The upper bound for the per-family error
    rate was set to $\mathtt{PFER}=1$. For the genomic data, the
    parameters were fixed to $\mathtt{B}=100$, $\mathtt{cutoff}=0.6$
    and $\mathtt{PFER}=10$.}  \fi
\item \textbf{Search space}: For computational reasons we chose to run
  some analyses chromosome by chromosome. Correction for multiple
  testing was done by controlling the false discovery rate
  \citep{BH:JRSS1995}. %
  {Since weak effects were expected, we also examined interactions
    with $p$-values $<0.05$ to discuss some aspects in relation with
    the phenotypes RTDBR and SNU.}  %
  
\end{itemize}

{Regarding the running time for the application, for about 2M SNPs and
  {329 bacterial genera}, the algorithm was able to perform the
  analysis in 250 min ($\sim$ 4 hours) with 10 CPU cores (Intel(R)
  Xeon(R) CPU E7-480 \verb+@+ 2.40GHz) and 2.5 Gb of memory.}

\subsection*{Results regarding Total Dry Biomass}

No significant interactions were found for this phenotype.

\subsection*{Results regarding the Root Total Dry Biomass Ratio}

{For RTDBR, four interactions were significant at $p$-value $<0.05$,
  distributed across three chromosomes, as shown in
  Table~\ref{tab:res_INRA}. The 365 210 SNPs allow recovering 9 007
  genes. A Gene Ontology enrichment analysis carried on the 4 490
  annotated genes identified ``hormone biosynthetic process'' (Fisher
  $p$-value of $2.10^{-17}$) or "antibiotic biosynthetic process"
  (Fisher $p$-value of $5.10^{-18}$), ``systemic acquired resistance''
  (Fisher $p$-value of $2.10^{-9}$) and ``cellular response to
  nitrogen starvation'' (Fisher $p$-value of $2.10^{-8}$) as four main
  overrepresented metabolic pathways involved in RTDBR variations under
  microbe interactions. The three first classes included almost
  redundant genes, mainly NBS-LRR kinase and 8 transcription
  factors. The fourth term ``cellular response to nitrogen
  starvation'' is composed mainly of lectin-domain receptor kinases
  genes also present in the three other classes and related to plant
  defense and of cysteine-rich receptor kinase genes, which are known
  to be regulated upon biotic and abiotic stress, such as salt and
  drought stress.  For the rhizosphere bacterial communities, 39
  genera were found in interaction with these genes. Also, 17, 9, and 6
  genera were affiliated to Proteobacteria, Actinobacteria, and
  Bacteroidtes respectively. Within Proteobacteria, 10 genera were
  identified as Alphaproteobacteria and 4 of them to the
  \textit{Rhizobiales} family, which is known to contribute to N
  nutrition of \textit{Medicago truncatula}. Plant disease resistance
  genes play a major role in the plant immune system that was induced
  during 
  pathogenic plant-microbial interactions but also during
  mutualistic plant-microbe
  interactions~\citep{hacquard2017interplay}. 
  None of the 39 bacterial genera identified was affiliated to genera
  known as plant pathogens. However, several of the bacterial genera
  identified were affiliated to genera known as plant symbiont or
  plant growth promoting bacteria. We could hypothesize that bacteria
  affiliated to these genera could be in positive interaction with the
  plant and induced some defense response.}

\subsection*{Results regarding Specific Nitrogen Uptake}

{For the SNU, we retrieved 157 698 significant SNPs and 5 476 genes
  from the three significant interactions, as shown in
  Table~\ref{tab:res_INRA}. Among the 3 136 annotated genes, the most
  over-represented biological process was the ``transmembrane receptor
  protein tyrosine kinase signalling pathway'' (Fisher $p$-value of
  $1.10^{-6}$), ``regulation of anion channel activity'' (Fisher
  $p$-value of $3.10^{-4}$) and ``lignin biosynthesis'' (Fisher
  $p$-value of $8.10^{-4}$). %
  The two first classes were partly redundant and mainly composed of
  LRR receptor kinase genes, known to be involved in plant innate
  immunity. The term ``regulation of anion channel activity'' was
  linked to other significant terms related to regulation to ion/anion
  transport. The ``lignin biosynthesis'' process included genes
  involved in lignin biosynthesis such as 8 caffeic acid
  O-methyltransferase genes, 3 cinnamyl alcohol dehydrogenase-like
  protein or 2 shikimate O-hydroxycinnamoyltransferase, which serve as
  building blocks in the formation of plant
  lignin~\citep{tu2010functional}. The colonization of plant host
  cells by bacteria involves the progressive remodeling of the
  plant–microbial interface for both \textit{Rhizobium}-Legume
  symbiosis~\citep{brewin2004plant} and pathogen
  bacteria~\citep{underwood2012plant}. In addition, the plant immune
  system is involved in symbiosis and during plant pathogen
  infections, and more generally with the plant
  microbiota~\citep{gourion2015rhizobium, hacquard2017interplay}. For
  the rhizosphere bacterial communities, 180 genera were found in
  interaction with these genes. 83, 31, 24 and 23 genera were
  affiliated to Proteobacteria, Firmicutes, Actinobacteria and
  Bacteroidetes respectively. In addition to the 13 genera belonging
  to the \textit{Rhizobiales} family, other OTUs were affiliated to
  bacteria genera harboring functional traits relating to the N cycle,
  such as nitrogen fixation, nitrate reduction to ammonium, and
  denitrification, which can contribute to plant nitrogen nutrition.
}

{Altogether, the mathematical method proposed here could support some
  biological hypothesis that need to be validated using other
  biological approaches combining plant mutant affected by these genes
  and simplified bacteria community defined on the genera identified.}

\begin{table}[h]
  \centering
  \begin{tabular}{c c c c c c c c}
    \hline
    \thead{PH} &  \thead{\#MG}
                      & \thead{CHR} & \thead{GP}
                      & \thead{\#SNPs}
                      & \thead{$p$-value} &
                                                  \thead{$q$-value}  \\\toprule
    RTDBR & 39 genera & 3 & 129:980206 & 6705 & 0.03 &	  0.18 \\
    RTDBR & 39 genera & 3 & 980235:32366703 & 196705 & 0.04 & 0.18 \\
    RTDBR & 39 genera & 7 & 21704918:33495621 & 68658 & 0.03 & 0.23 \\
    RTDBR & 39 genera & 8  & 50:18024047 & 93142 & 0.02 & 0.14 \\
    SNU & 180 genera & 2 & 38539843:45729381 & 33033 & 0.04 & 0.13 \\
    SNU & 180 genera & 6 & 33985403:35275305 & 6174 & 0.04 & 0.13 \\
    SNU & 180 genera & 8 & 18024755:45569421 & 156827 & 0.05 & 0.09 \\ 
    \bottomrule
  \end{tabular}
  \smallskip
  \caption{Results of the search for interactions using the
    $\rho$-SICOMORE method. From left to right, the names of the
    columns are: PH for the phenotype studied; \#MG for the number of
    genera; CHR for the chromosome; GP for the genomic postion (pb)
    and \#SNPs for the number of SNPs in the genomic region.}
\label{tab:res_INRA}
\end{table}

\section{Conclusion}
\label{sec:conclusion}

\subsection*{Synthesis} The detection of interaction effects in a
high-dimensional setting remains a difficult problem because multiple
testing is onerous and because effects are small in terms of their
significance.  {In this work, we proposed SICOMORE, a method that
  reduces the dimension of the search space by selecting a subset of
  compressed variables obtained from the biological characteristics of
  complementary datasets.}

Our approach has demonstrated its ability to uncover interaction
effects with a high statistical power. In our simulations, where
sample sizes, noise, and the number of true interactions all varied,
SICOMORE always exhibited stronger recall than both MLGL and
glinternet.  SICOMORE combines the strengths of different methods in a
powerful single algorithm.  SICOMORE is also significantly more
efficient than the others in terms of computation time.

SICOMORE was able to detect interactions between the genome of
\textit{Medicago truncatula} and its rhizosphere, which are linked to
the Root Total Biomass Ratio as well as its Specific Nitrogen Uptake.

\subsection*{{Extensions}}

{Although our approach as presented here concerns the detection of
  interactions between genomic and metagenomic markers, it should be
  noted that two major extensions are available.
  \begin{enumerate}
  \item SICOMORE can be applied to any kind of numerical data, as long
    as an underlying hierarchical or group structure is available
    (such as a correlation structure, for instance). In particular, our
    method can handle shotgun sequencing as well as other omics data,
    or even clinical follow-up, which often takes the form of categorical
    data that can be easily structured.
  \item The compression scheme used in SICOMORE means that the model
    of interactions can easily be extended to $V > 2$ different
    datasets.  This opens the way to tackling a variety of other
    problems where different sources of information may be utilized,
    such as in precision medicine, for instance.
  \end{enumerate}
  The \texttt{R} package already incorporates these two possibilities.}

\subsection*{Perspectives}

\ifstabs Given these interesting results {and possible extensions,
  there are other aspects that may be interesting to address in future
  works, with a view to improving SICOMORE further in terms of model
  consistency.  Although the Lasso procedure is relevant for dimension
  reduction purposes, it may induce some biases in the multiple
  testing procedure used afterwards, since the variable selection step
  is performed before the $p$-values are adjusted. One way around this
  problem might be to use post-hoc inference for multiple comparisons
  \citep{goeman2011multiple}. These kinds of extensions should have a
  positive impact on precision results.%
  \else%
  Given these interesting results and possible extensions, there are
  aspects that it might be interesting to address in future works,
  with a view to improving SICOMORE further in terms of model
  consistency.  First, the variable selection step for selecting the
  supervariables in the two complementary datasets can be subject to
  instability when setting the amount of selection. This might be
  addressed using resampling techniques \citep{Bach:ICML08,
    MB:JRSSB10}. Second, although the Lasso procedure is relevant for
  dimension reduction purposes, it may induce some biases in the
  multiple testing procedure used afterwards, since the variable
  selection step is performed before the $p$-values are adjusted. One
  way around this problem might be to use post-hoc inference for
  multiple comparisons \citep{goeman2011multiple}. These kinds of
  extensions should have a positive impact on precision results.%
 \fi

\bibliographystyle{plainnat}
\bibliography{biblio}

\newpage
\appendix

\section{Supplementary results}
\label{sec:supp}

\begin{figure}[htb]
  \centering
  \subfigure[$I=1$]{
    \includegraphics[width=0.92\textwidth]{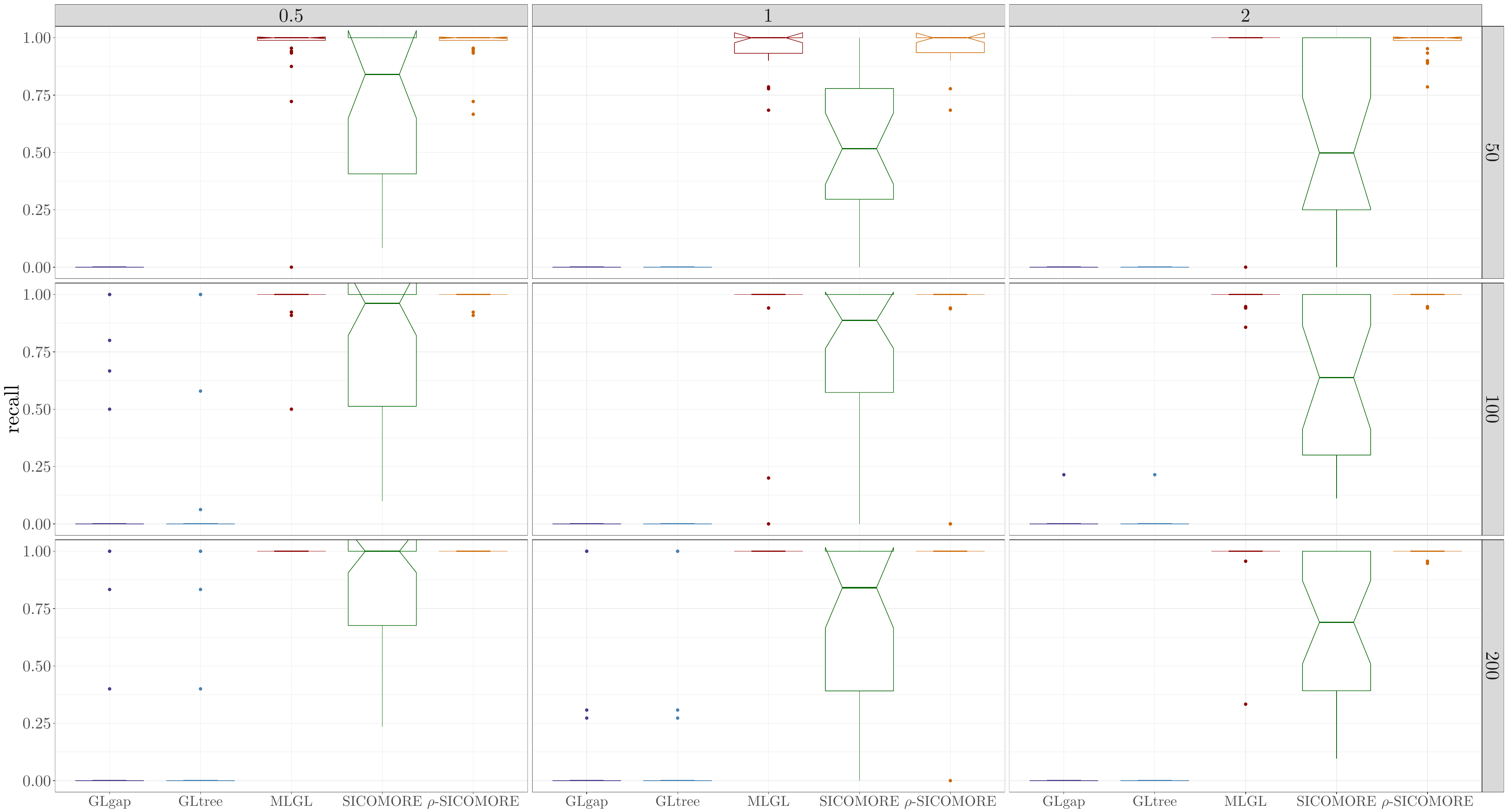}
    \label{fig:Recall_1}
  }\\
  \subfigure[$I=3$]{
    \includegraphics[width=0.92\textwidth]{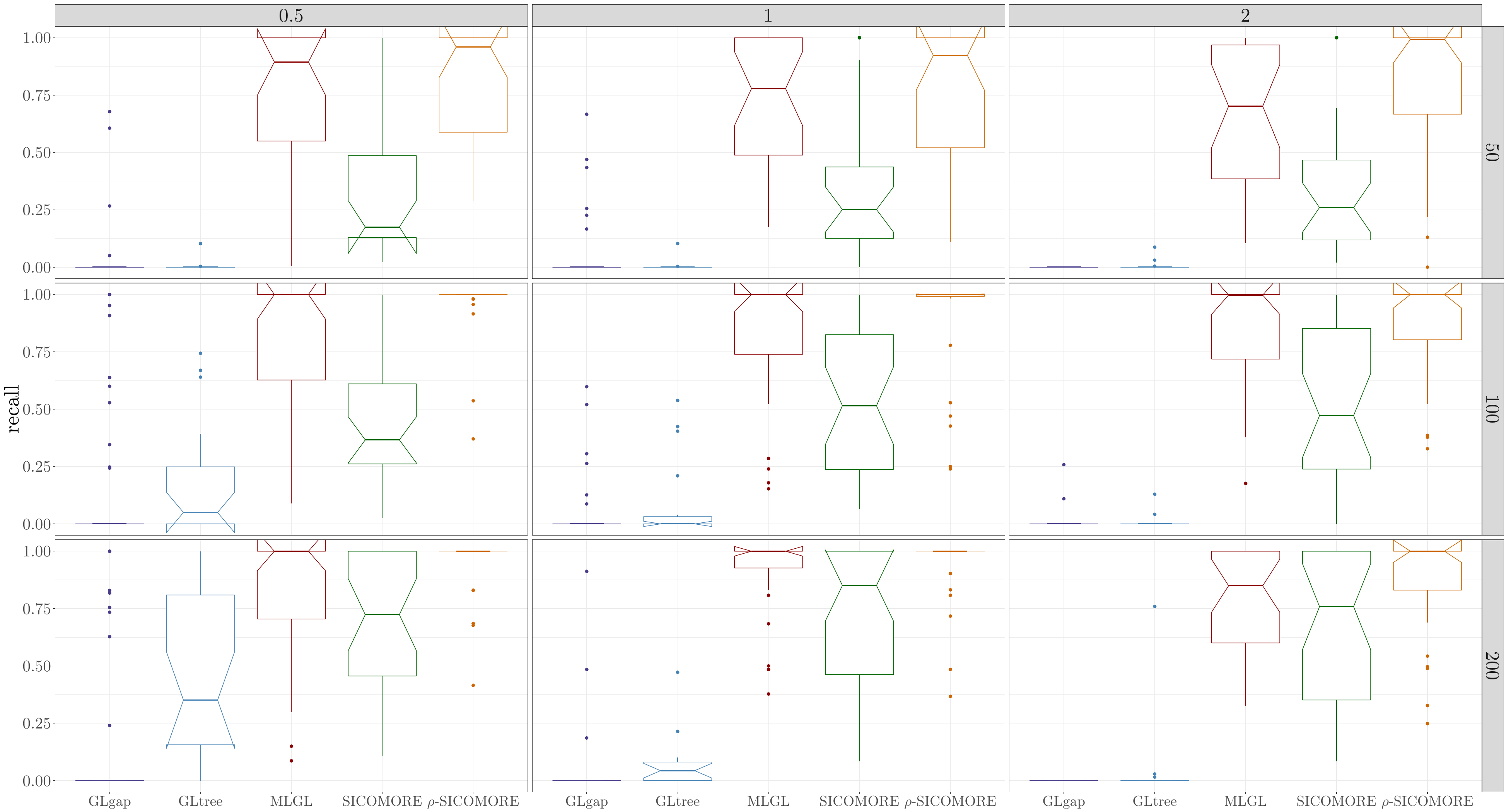}
    \label{fig:Recall_3}
  }
\end{figure}
\begin{figure}[htb]
  \subfigure[$I=5$]{
    \includegraphics[width=0.92\textwidth]{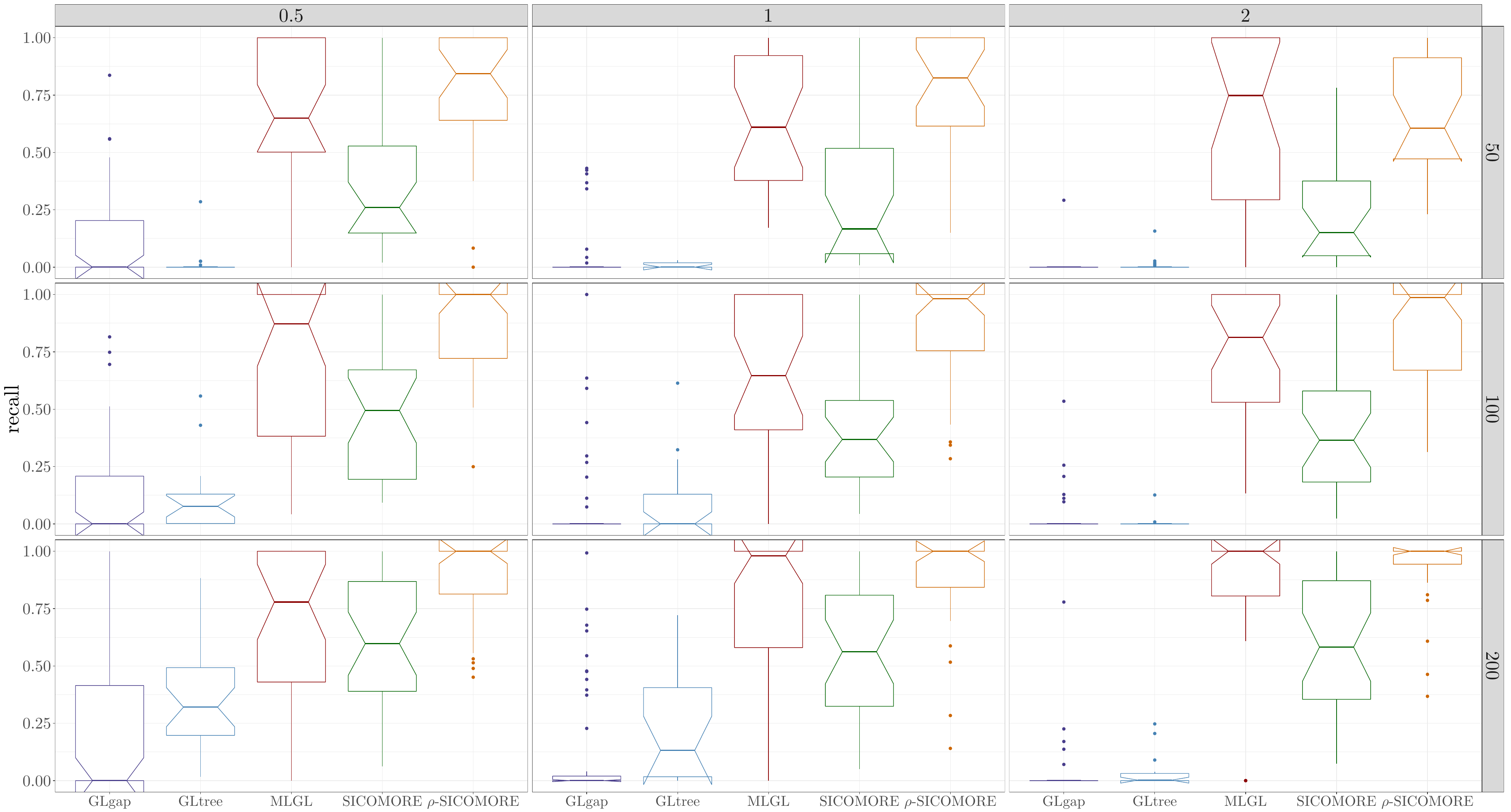}
    \label{fig:Recall_5}
  }\\
  \subfigure[$I=10$]{
    \includegraphics[width=0.92\textwidth]{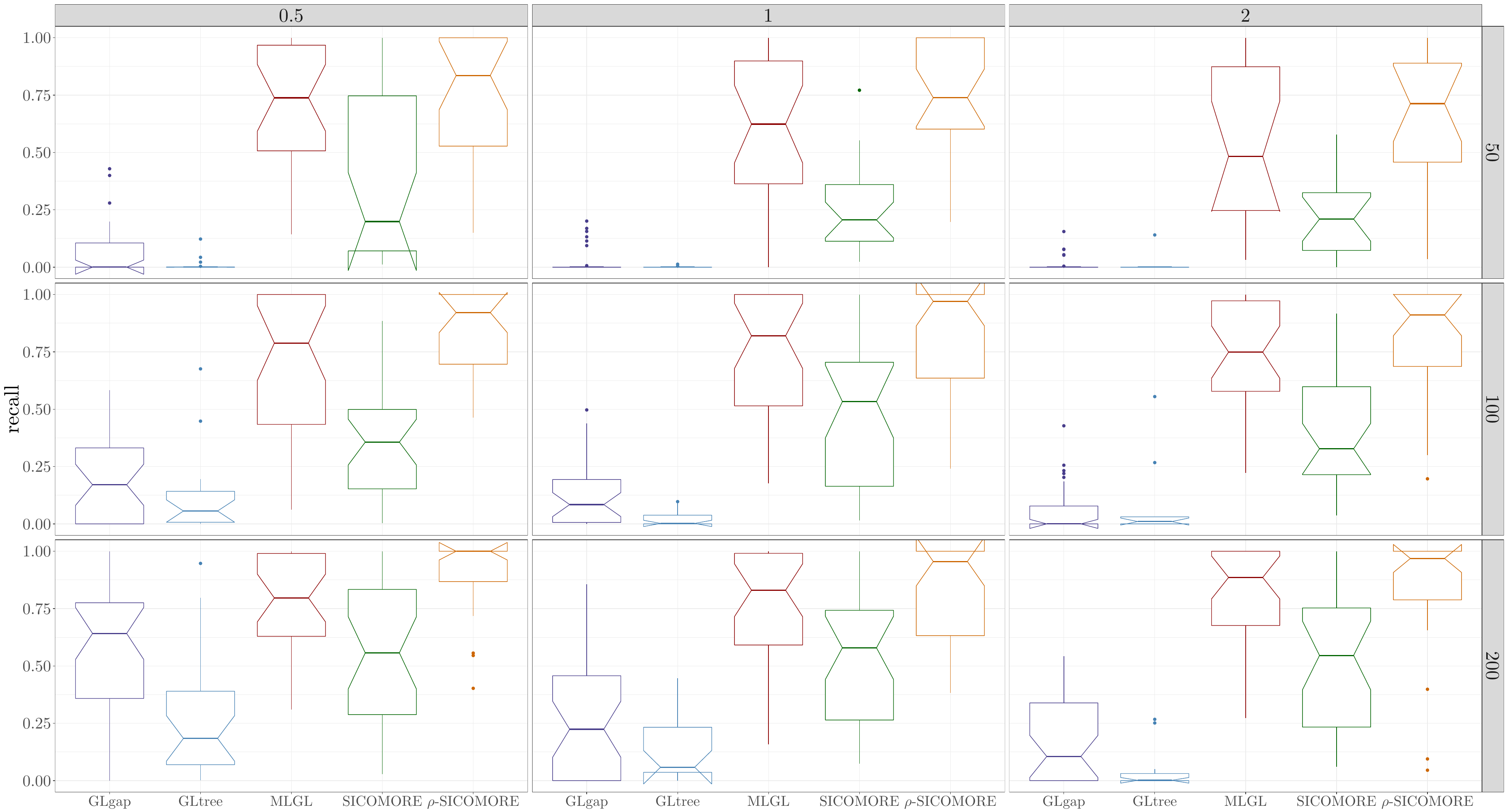}
    \label{fig:Recall_10}
  }
  \caption{Boxplots for Recall obtained on the numerical simulations
    with a Bonferroni-Holm correction for $I=\{1, 3, 5, 10\}$
    blocs. The lines show the results for different number of
    observations (top: $N=50$, middle: $N=100$ and bottom: $N=200$)
    and the columns the difficulty of the problem (left:
    $\epsilon=0.5$, middle: $\epsilon=1$ and right: $\epsilon=2$). The
    boxplots are best seen in colors: from the left to the right,
    GLgap is in purple, GLtree is in blue, MLGL is in red, SICOMORE is
    in green, $\rho$-SICOMORE is in orange.}
  \label{fig:Recall_13510}
\end{figure}

\begin{figure}[htb]
  \centering
  \subfigure[$I=1$]{
    \label{fig:Precisions_1}
    \includegraphics[width=0.92\textwidth]{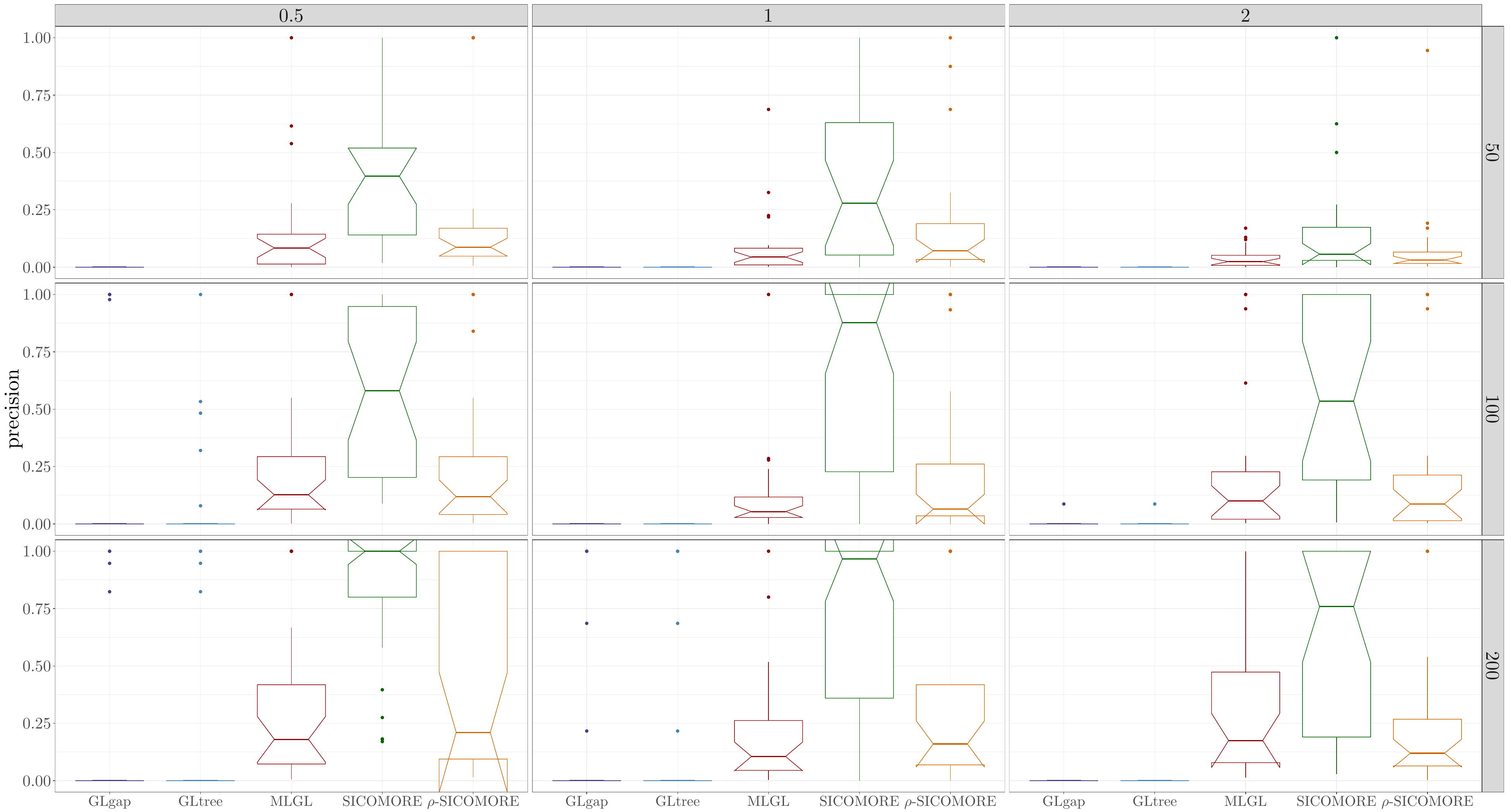}
  }\\
  \subfigure[$I=3$]{
    \label{fig:Precisions_3}
    \includegraphics[width=0.92\textwidth]{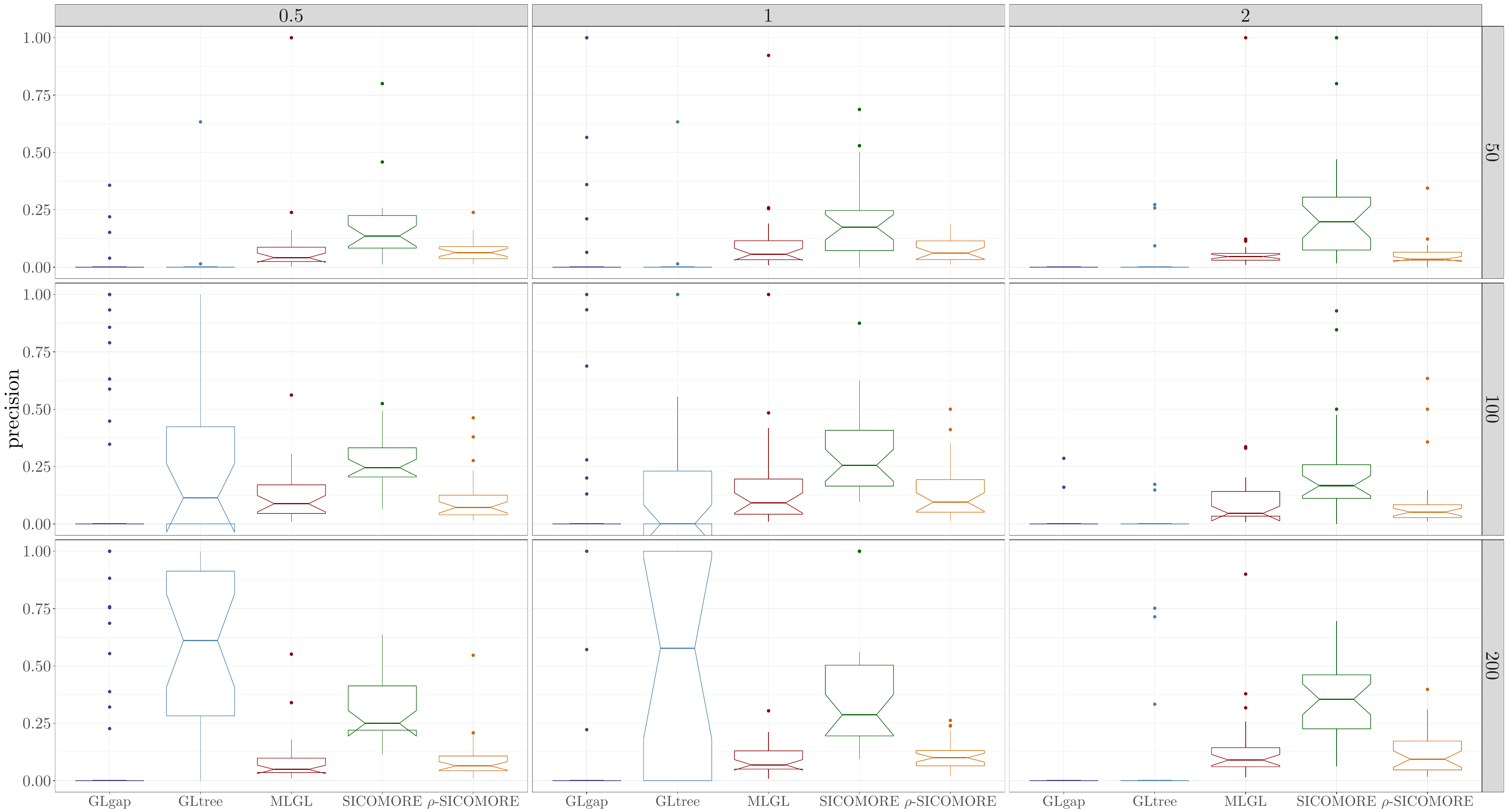}
  }
\end{figure}
\begin{figure}[htb]
  \subfigure[$I=5$]{
    \label{fig:Precisions_5}
    \includegraphics[width=0.92\textwidth]{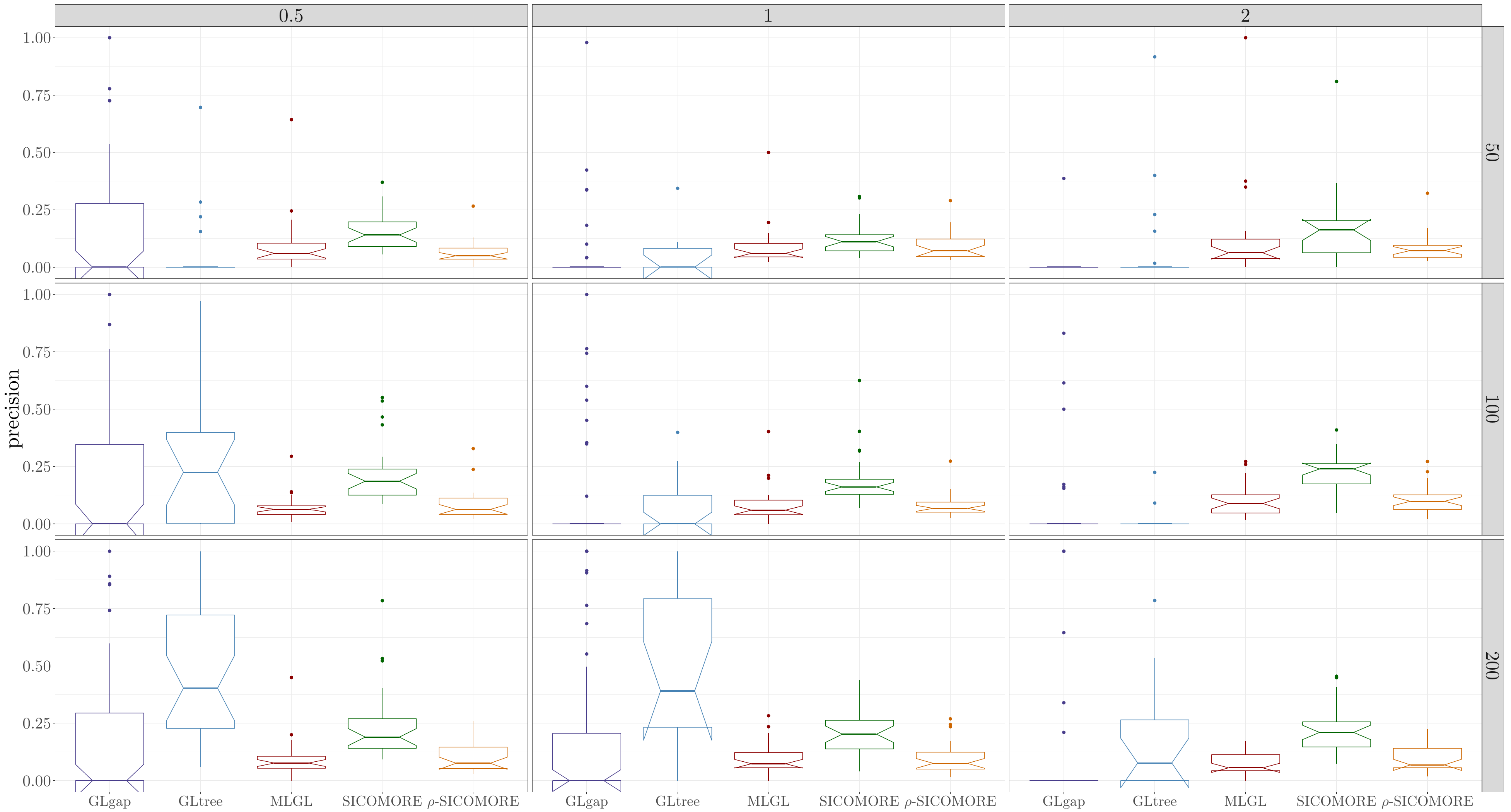}
  }\\
  \subfigure[$I=10$]{
    \label{fig:Precisions_10}
    \includegraphics[width=0.92\textwidth]{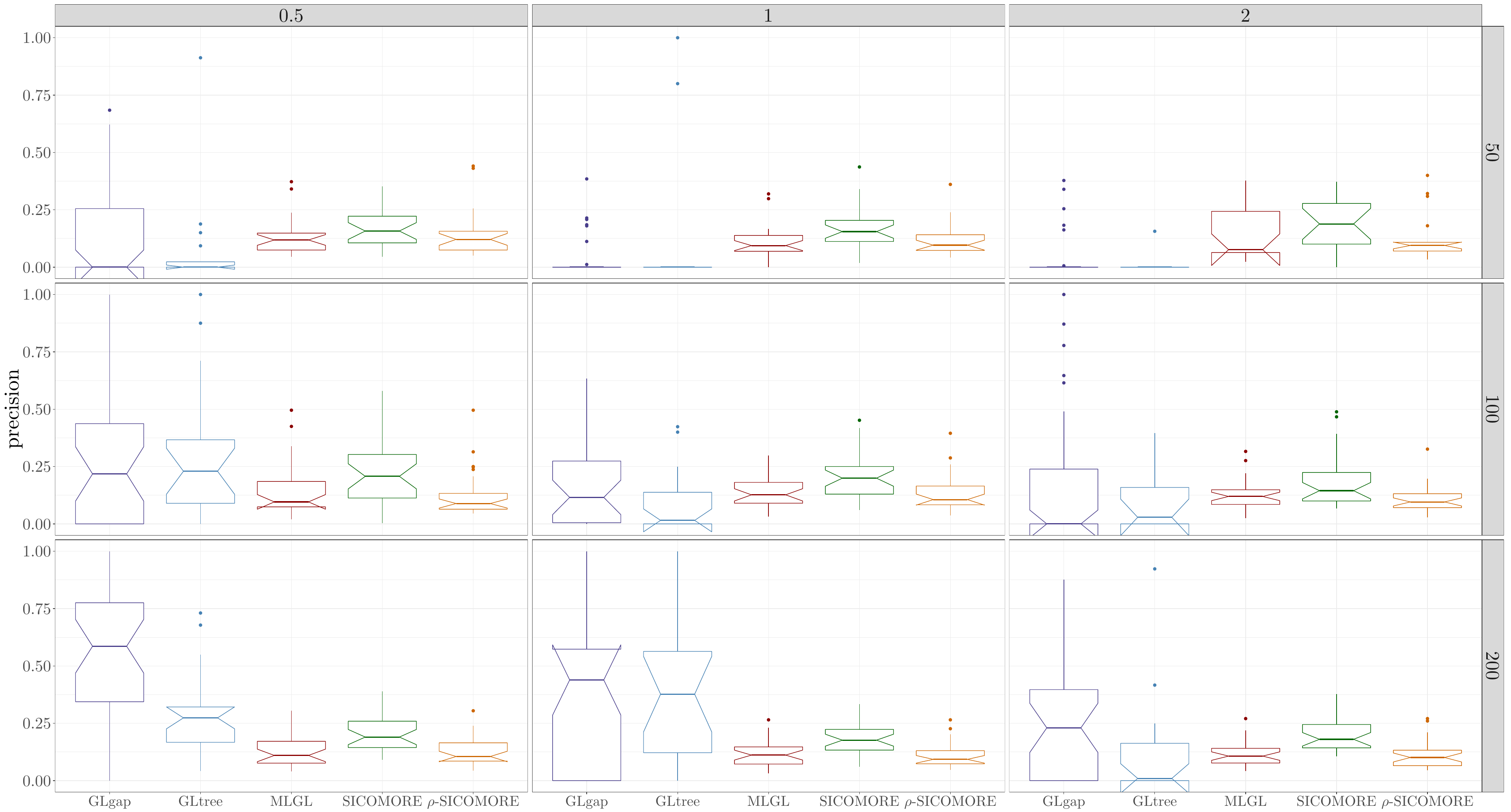}
  }
  \caption{Boxplots for Precision obtained on the numerical
    simulations with a Bonferroni-Holm correction for
    $I=\{1, 3, 5, 10\}$ blocs. The lines show the results for
    different number of observations (top: $N=50$, middle: $N=100$ and
    bottom: $N=200$) and the columns the difficulty of the problem
    (left: $\epsilon=0.5$, middle: $\epsilon=1$ and right:
    $\epsilon=2$). The boxplots are best seen in colors: from the left
    to the right, GLgap is in purple, GLtree is in blue, MLGL is in
    red, SICOMORE is in green, $\rho$-SICOMORE is in orange.}
  \label{fig:Precisions_13510}
\end{figure}
\FloatBarrier

\section{Supplementary data for the \textit{Medicago truncatula}
  example} 
\label{sec:supp_data}

With this supplemental data, we intend to provide details to explain
the metabarcoding analysis leading to the OTUs used in the
\textit{Medicago truncatula} example. This is a part of a side
research paper in preparation of Anouk Zancarini, Christine Le Signor
and Christophe Mougel.

\subsection*{Metabarcoding analysis}

To assess the bacterial communities, the variable region V4 of the 16S
rRNA gene was amplified using the 479F and 888R primers and sequenced
using Illumina MiSeq sequencing technology (paired-end 2$\times$250
pb). Bioinformatic analyses were done using the GnS-PIPE developed by
the GenoSol platform (INRA, Dijon,
France)~\citep{terrat2012molecular}.
The details of all steps have been already described
previously~\citep{terrat2015meta}.

After preprocessing, alignment and clustering of reads at 95\% of
similarity, a filtering step was carried out to check all
single-singletons~\footnote{Single-singletons are reads detected only
  once and not clustered.} to eliminate PCR chimeras and large
sequencing errors produced by the PCR step, based on the quality of
their taxonomic assignments. More precisely, each single-singleton was
compared with a dedicated reference database from the Silva curated
database using similarity approaches (USEARCH), with sequences longer
than 500 nucleotides, and kept only if their identity was higher than
the defined threshold (95\%). The number of high-quality reads for
each sample~\footnote{There are 10 000 high-quality reads for each
  sample.}  was normalized by random selection to allow an efficient
comparison of the datasets and to avoid biased community comparisons.

Then, as the analysis of microbial community richness relies on the
construction of similarity clusters (called OTUs), we chose here to
use OTUs to examine the distribution of 16S rRNA gene sequences in our
datasets. This clustering was realized with a Perl script program that
groups rare reads to abundant ones, and does not count the differences
in homopolymer lengths. Finally, the global contingency table of OTUs
was obtained with the samples in lines and the OTUs in columns,
indicating the number of reads in each OTU for all samples. The
taxonomy of each OTU was determined based on the taxonomy of all reads
encompassed in the OTU. More precisely, 
an OTU composed of more than 90\% of reads of a given taxonomy is
assigned to this taxonomy.  The community structure was then
characterized using weighted UniFrac
distance~\citep{lozupone2005unifrac}
calculated
with the PycoGent package~\citep{knight2007pycogent} 
on a phylogenetic tree computed using FastTree and the most abundant
sequence to represent each OTU. 

One sample was removed because of its too
low-depth~\citep{weiss2017normalization}. 
The OTUs with counts lower than 41 over all the samples were
filtered. The threshold of 41 was determined using the following
procedure: 
for a threshold varying from 1 to 150, we calculated the number of
OTUs whose total counts over all the samples is below this threshold. 
The selected threshold is the one for which the number of OTUs does
not increase when the threshold increases by one.  Then, the number of
reads in each OTU was first summed for the three replicates for each
plant genotype and a between-sample normalization was performed in
order to correct for the different sequencing depth. Each sample was
scaled by a size factor calculated as the ratio between the total
number of counts in this sample and the mean of total counts across
all samples. Finally, for each plant genotype, the number of reads
were summed for OTU belonging to the same genus. OTUs that had unknown
taxonomic assignment at genus level were discarded. Thus, a total of
155 samples and 329 genus were finally analysed.


All raw data sets are publicly available in the European Nucleotide
Archive (ENA) of EMBL-EBI database system under project accession
PRJEB25849 entitled "\emph{Genome-wide association study of
  \textit{Medicago truncatula} rhizosphere microbial communities and
  plant nutritional strategies}" with raw sequences accession
(ERR2495157 to ERR2495714).

\subsection*{Taxonomic affiliations of OTUs}

\begin{figure}[h]
  \begin{center}
    \includegraphics[scale=.65]{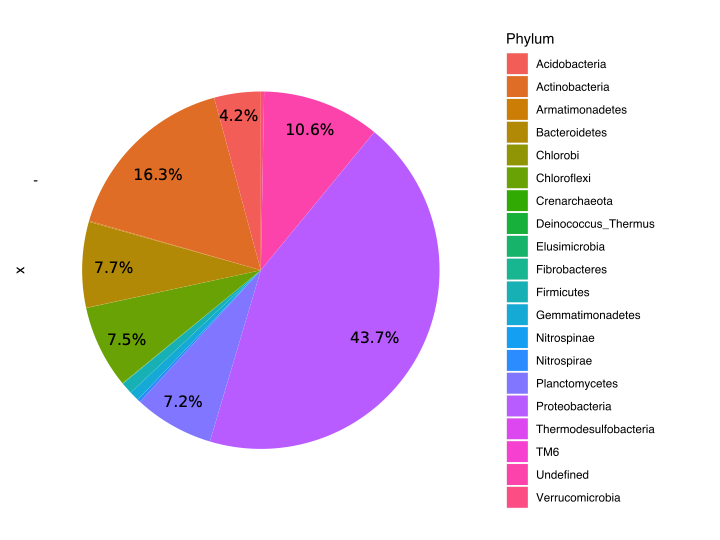}
  \end{center}
  \caption{Taxonomic affiliation of the OTUs at Phylum level.}
  \label{fig:OTU}
\end{figure}

We provide a pie chart that depicts the taxonomic affiliation of the
OTUs at phylum level in Figure~\ref{fig:OTU}. This results will be
presented as a boxplot and discussed in the side paper still in
preparation.

\end{document}

%% file: commandes-math.tex
\newcommand{\IR}{\mathbb{R}}

\newcommand{\bx}{\mathbf{x}}
\newcommand{\tbx}{\tilde{\mathbf{x}}}
\newcommand{\bX}{\mathbf{X}}

\newcommand{\bbeta}{\boldsymbol{\beta}}

\newcommand{\btheta}{\boldsymbol{\theta}}

\newcommand{\bepsilon}{\boldsymbol{\epsilon}}
\newcommand{\bgamma}{\boldsymbol{\gamma}}

\newcommand{\bphi}{\boldsymbol{\phi}}

\newcommand{\bSigma}{\boldsymbol{\Sigma}}

\newcommand{\bDelta}{\boldsymbol{\Delta}}

\newcommand{\vX}{\mathrm{X}}
\newcommand{\vZ}{\mathrm{Z}}

\DeclareMathOperator*{\argmin}{argmin}

\newcommand{\vd}{\mathit{M}}
\newcommand{\ivd}{m}

\newcommand{\vu}{\mathit{G}}
\newcommand{\ivu}{g}

\newcommand{\gvu}{\mathcal{G}}

\newcommand{\gvd}{\mathcal{M}}

\theoremstyle{remark}
\newtheorem*{rem}{Remark}



%% file: commandes-dessin.tex
\tikzstyle{model}=[circle,inner sep=0pt,minimum size=2mm]
\tikzstyle{group}=[circle,draw=red!50,fill=red!20,thick,
                   inner sep=0pt,minimum size=6mm]
\tikzstyle{rgroup}=[circle,draw=red!80,fill=red!50,thick,
                   inner sep=0pt,minimum size=5mm]
\tikzstyle{leaf}=[rectangle,draw=blue!50,fill=blue!20,thick,
                        inner sep=0pt,minimum width=8mm, minimum
                   height=4mm]

\tikzstyle{l0}=[circle,draw=rootcolor,fill=rootcolor!35!white,thick,
                   inner sep=0pt,minimum size=1.25mm, line width =
                   0.75pt]
\tikzstyle{l1}=[circle,draw=blue,fill=blue!35!white,thick,
                   inner sep=0pt,minimum size=1.25mm, line width =
                   0.75pt]
\tikzstyle{l2}=[circle,draw=red,fill=red!35!white,thick,
                   inner sep=0pt,minimum size=1.25mm, line width =
                   0.75pt]
\tikzstyle{l3}=[circle,draw=violet,fill=violet!35!white,thick,
                   inner sep=0pt,minimum size=1.25mm, line width =
                   0.75pt]

\definecolor{bgfig}{gray}{1}

\makeatletter 

\pgfkeys{/pgf/decoration/.cd, 
  start color/.store in=\startcolor, 
  end color/.store in=\endcolor 
} 

\pgfdeclaredecoration{color change arrow}{initial}{ 
  \state{initial}[width=0pt, next state=line, persistent precomputation={%
    \pgfmathdivide{50}{\pgfdecoratedpathlength}%
    \let\increment=\pgfmathresult%
    \def\x{0}%
  }]{} 
  \state{line}[width=0.5pt, switch if less than=5.0\pgflinewidth to final, 
    persistent postcomputation={%
      \pgfmathadd@{\x}{\increment}%
      \let\x=\pgfmathresult%
    }]{%
    \pgfsetlinewidth{\pgflinewidth}%
    \pgfsetarrows{-}%
    \pgfpathmoveto{\pgfpointorigin}%
    \pgfpathlineto{\pgfqpoint{0.75pt}{0pt}}%
    \pgfsetstrokecolor{\endcolor!\x!\startcolor}%
    \pgfusepath{stroke}%
  } 
  \state{final}{%
    \pgfsetlinewidth{\pgflinewidth}%
    \pgfpathmoveto{\pgfpointorigin}%
    \pgfpathlineto{\pgfpointdecoratedpathlast\advance\pgf@x by2.0\pgflinewidth}%
    \color{\endcolor!\x!\startcolor}%
    \pgfusepath{stroke}%
  } 
} 

\makeatother

%% file: commandes-colors.tex
\definecolor{ensiievert}{RGB}{93, 176, 0}
\definecolor{darkgreen}{RGB}{0, 102, 51}
\definecolor{rootcolor}{RGB}{0, 100, 0}
\definecolor{vdcolor}{RGB}{76, 217, 100}
\definecolor{vucolor}{RGB}{255, 149, 0}

\definecolor{ensiiegris}{RGB}{67, 75, 76}
\definecolor{bgfig}{gray}{1}
\definecolor{examen}{RGB}{200, 15, 15}